\documentclass[twocolumn]{aastex63}
\usepackage{graphicx}
\usepackage{subfigure}
\usepackage{color, hyperref, epsfig}
\usepackage{apjfonts, natbib}
\usepackage{appendix}
\usepackage{float}
\usepackage{bm}
\usepackage{lineno}

\newcommand{\ha}{H\ensuremath{\alpha}}

\newcommand{\lum}{erg\,s\ensuremath{^{-1}}}

\newcommand{\flux}{erg\,s$^{-1}$\,cm$^{-2}$}

\newcommand{\msun}{\ensuremath{M_{\odot}}}
\newcommand{\lsun}{\ensuremath{L_{\odot}}}
\newcommand{\kms}{\ensuremath{\mathrm{km\,s^{-1}}}}

\newcommand{\wise}{\emph{WISE}}
\newcommand{\neowise}{\emph{NEOWISE}}

\newcommand{\nh}{\ensuremath{N\mathrm{_{H}}}}

\shorttitle{Superluminous Supernova 2017egm}
\shortauthors{Zhu et al.}

\begin{document}

\title{\Large SN~2017egm: A Helium-rich Superluminous Supernova with Multiple Bumps in the Light Curves}
\correspondingauthor{Subo Dong, Ning Jiang}
\email{dongsubo@pku.edu.cn, jnac@ustc.edu.cn}
\author[0000-0003-3824-9496]{Jiazheng Zhu}
\affiliation{CAS Key laboratory for Research in Galaxies and Cosmology,
Department of Astronomy, University of Science and Technology of China, 
Hefei, 230026, China; jiazheng@mail.ustc.edu.cn, jnac@ustc.edu.cn}
\affiliation{School of Astronomy and Space Sciences,
University of Science and Technology of China, Hefei, 230026, China}
\author[0000-0002-7152-3621]{Ning Jiang}
\affiliation{CAS Key laboratory for Research in Galaxies and Cosmology,
Department of Astronomy, University of Science and Technology of China, 
Hefei, 230026, China; jiazheng@mail.ustc.edu.cn, jnac@ustc.edu.cn}
\affiliation{School of Astronomy and Space Sciences,
University of Science and Technology of China, Hefei, 230026, China}
\author[0000-0002-1027-0990]{Subo Dong}
\affiliation{Kavli Institute for Astronomy and Astrophysics, Peking University, Yi He Yuan Road 5, Hai Dian District, Beijing 100871, People's Republic of China; dongsubo@pku.edu.cn}
\author[0000-0003-3460-0103]{Alexei V. Filippenko} \affil{Department of Astronomy, University of California, Berkeley, CA 94720-3411, USA} 
\author[0000-0003-3096-759X]{Richard J.~Rudy} \affil{Physical Sciences Laboratory, The Aerospace Corporation M2-266, P.O. Box 92957, Los Angeles, CA 90009, USA}\affil{Kookoosint Scientific, 1530 Calle Portada, Camarillo, CA 93010, USA}
\author[0000-0002-7259-4624]{A.~Pastorello}\affil{INAF-Osservatorio Astronomico di Padova, Vicolo dell'Osservatorio 5, I-35122 Padova, Italy}
\author[0000-0002-5221-7557]{Christopher Ashall} \affil{Department of Physics, Virginia Tech, Blacksburg, VA 24061, USA}
\author[0000-0003-3529-3854]{Subhash Bose}\affil{Department of Astronomy, The Ohio State University, 140 W. 18th Avenue, Columbus, OH 43210, USA} \affil{Center for Cosmology and AstroParticle Physics (CCAPP), The Ohio State University, 191 W. Woodruff Avenue, Columbus, OH 43210, USA}
\author{R.~S. Post} \affil{Post Observatory, Lexington, MA 02421, USA}
\author[0000-0001-7485-3020]{D.~Bersier}\affil{Astrophysics Research Institute, Liverpool John Moores University, 146 Brownlow Hill, Liverpool L3 5RF, UK}
\author[0000-0002-3256-0016]{Stefano Benetti} \affil{INAF-Osservatorio Astronomico di Padova, Vicolo dell'Osservatorio 5, I-35122 Padova, Italy}	
\author[0000-0001-5955-2502]{Thomas G.~Brink} \affil{Department of Astronomy, University of California, Berkeley, CA 94720-3411, USA}
\author[0000-0003-0853-6427]{Ping Chen} \affil{Department of Particle Physics and Astrophysics Weizmann Institute of Science 234 Herzl St. 7610001 Rehovot, Israel}
\author[0000-0002-4757-8622]{Liming Dou}\affiliation{Department of Astronomy, Guangzhou University, Guangzhou 510006, China}	
\author[0000-0002-1381-9125]{N.~Elias-Rosa} \affil{INAF-Osservatorio Astronomico di Padova, Vicolo dell'Osservatorio 5, I-35122 Padova, Italy}	
\author[0000-0002-3664-8082]{Peter Lundqvist} \affil{Department of Astronomy and The Oskar Klein Centre, AlbaNova University Center, Stockholm University, SE-10691 Stockholm, Sweden}
\author[0000-0001-7497-2994]{Seppo Mattila} \affil{Tuorla Observatory, Department of Physics and Astronomy, FI-20014 University of Turku, Finland}
\affil{School of Sciences, European University Cyprus, Diogenes street, Engomi, 1516 Nicosia, Cyprus}
\author{Ray W.~Russell}\affil{Physical Sciences Laboratory, The Aerospace Corporation M2-266, P.O. Box 92957, Los Angeles, CA 90009, USA}
\author[0000-0003-1799-1755]{Michael L.~Sitko}\affil{Center for Exoplanetary Systems, Space Science Institute, 4750 Walnut Street, Suite 205, Boulder, CO 80301, USA}
\author[0000-0001-6566-9192]{Auni Somero} \affil{Tuorla Observatory, Department of Physics and Astronomy, FI-20014 University of Turku, Finland}	
\author[0000-0002-5571-1833]{M.~D.~Stritzinger} \affil{Department of Physics and Astronomy, Aarhus University, Ny Munkegade 120, DK-8000 Aarhus C, Denmark}	
\author[0000-0002-1517-6792]{Tinggui Wang}
\affiliation{CAS Key laboratory for Research in Galaxies and Cosmology,
Department of Astronomy, University of Science and Technology of China, Hefei, 230026, China; jiazheng@mail.ustc.edu.cn, jnac@ustc.edu.cn}
\affiliation{School of Astronomy and Space Sciences,
University of Science and Technology of China, Hefei, 230026, China}								
\author[0000-0001-6272-5507]{Peter J.~Brown} \affil{George P. and Cynthia Woods Mitchell Institute for Fundamental Physics \& Astronomy, Texas A. \& M. University, Department of Physics and Astronomy, 4242 TAMU, College Station, TX 77843, USA}
\author[0000-0001-5008-8619]{E.~Cappellaro} \affil{INAF-Osservatorio Astronomico di Padova, Vicolo dell'Osservatorio 5, I-35122 Padova, Italy}					
\author[0000-0003-2191-1674]{Morgan Fraser} \affil{School of Physics, O'Brien Centre for Science North, University College Dublin, Belfield, Dublin 4}	
\author[0000-0001-8257-3512]{Erkki Kankare}\affil{Tuorla Observatory, Department of Physics and Astronomy, University of Turku, FI-20014 Turku, Finland}
\author{S.~Moran}\affil{Tuorla Observatory, Department of Physics and Astronomy, FI-20014 University of Turku, Finland}	
\author[0000-0003-0486-6242]{Simon Prentice} \affil{Astrophysics Research Institute, Liverpool John Moores University, Liverpool, L3 5RF, UK}
\author[0000-0002-5578-9219]{Tapio Pursimo} \affil{Nordic Optical Telescope, Apartado 474, 38700 Santa Cruz de La Palma, Spain}
\author[0000-0002-1022-6463]{T.~M.~Reynolds} \affil{Cosmic Dawn Center (DAWN), Niels Bohr Institute, University of Copenhagen, Jagtvej 128, 2200 København N, Denmark} \affil{Tuorla Observatory, Department of Physics and Astronomy, FI-20014 University of Turku, Finland}
\author[0000-0002-2636-6508]{WeiKang Zheng} \affil{Department of Astronomy, University of California, Berkeley, CA 94720-3411, USA}

\begin{abstract}

When discovered, SN~2017egm was the closest (redshift $z=0.03$) hydrogen-poor superluminous supernova (SLSN-I) and a rare case that exploded in a massive and metal-rich galaxy. Thus, it has since been extensively observed and studied. We report spectroscopic data showing strong emission at around He~I $\lambda$10,830 and four He~I absorption lines in the optical. Consequently, we classify SN~2017egm as a member of an emerging population of helium-rich SLSNe-I (i.e., SLSNe-Ib). We also present our late-time photometric observations. By combining them with archival data, we analyze high-cadence ultra-violet, optical, and near-infrared light curves spanning from early pre-peak ($\sim -20$\,d) to late phases ($\sim +300$\,d). We obtain its most complete bolometric light curve, in which multiple bumps are identified. None of the previously proposed models can satisfactorily explain all main light-curve features, while multiple interactions between the ejecta and circumstellar material (CSM) may explain the undulating features. The prominent infrared excess with a blackbody luminosity of $10^7$--$10^8\,\lsun$ detected in SN~2017egm could originate from the emission of either an echo of a pre-existing dust shell, or newly-formed dust, offering an additional piece of evidence supporting the ejecta-CSM interaction model. Moreover, our analysis of deep $Chandra$ observations yields the tightest-ever constraint on the X-ray emission of an SLSN-I, amounting to an X-ray-to-optical luminosity ratio $\lesssim 10^{-3}$ at late phases ($\sim100$--200\,d), which could help explore its close environment and central engine.

\end{abstract}


\section{Introduction}

Thanks to the rise of wide-field time-domain surveys, a rare class of supernovae (SNe) which are $\sim$2--3 mag more luminous than normal SNe have been revealed in the past one and a half decades, known as the superluminous supernovae (SLSNe; see a recent review by \citealt{galyam2019}). As with normal SNe \citep{Filippenko97}, SLSNe are categorized into two classes based on their spectra around maximum light: hydrogen-poor (SLSNe-I) and hydrogen-rich events (SLSNe-II) \citep{galyam2019}.

Since the discovery of the first SLSN-I over 15\,yr ago (SN 2005ap; \citealt{Quimby2007}), more than 100 such events have been found, and identifications of diverse properties among them have motivated subclassifications. Subclasses with rapidly and slowly evolving light curves have been proposed  \citep{galyam2012, Inserra2018, Quimby2018}, but whether clear divisions can be made are debated (e.g., \citealt{Nicholl2017b, Lunnan2018}). Spectroscopically, their early-time spectra often show a characteristic W-shaped feature near 4200\,\AA\ produced by a pair of broad absorption features associated with O~II (\citealt{Quimby2011}). SLSNe-I has long been linked to SNe-Ic due to their similarities in post-peak spectra and a lack of both H and He \citep{Pastorello10, Quimby2011}. During the past few years, several SNe initially classified as SLSN-I have been found with He spectral features.  PTF10hgi shows both H and He lines \citep{Quimby2018}, which is followed by a small sample of He-rich SLSNe-I classified as SLSNe-Ib from the Zwicky Transient Facility (ZTF; \citealt{Yan2020}). However, the studies of SLSNe-Ib still lack high-quality multiwavelength photometric and spectroscopic observations covering the early phase to the late stages. 

The powering mechanisms of SLSNe-I remain elusive \citep{Moriya2018, galyam2019}. Several possible mechanisms beyond the conventional means such as radioactive decay of $^{56}$Ni have been proposed to explain their extreme radiative power, including energy injection from a rapidly-rotating, highly magnetic neutron star (e.g., \citealt{Kasen2010}, \citealt{Woosley2010}), the interaction of the SN ejecta with circumstellar material (CSM; e.g., \citealt{Chevalier2011}, \citealt{Chatzopoulos2013}, \citealt{Sorokina2016}), and the pair-instability mechanism (e.g., \citealt{Heger2002}; \citealt{galyam2009}). Magnetar models are often invoked in fitting the light curves. However, some observational evidence already indicates that the magnetar model cannot be the only process affecting the luminosity and morphology for SLSNe-I, and multiple processes may contribute to the optical emission. For example, the light-curve undulations and sometimes prominent post-peak bumps seen in some SLSNe-I light curves can not be explained by a simple magnetar model (e.g., \citealt{Nicholl2016}; \citealt{Yan2017a, Yan2017b}; \citealt{Anderson2018}; \citealt{Lunnan2020}). The CSM interaction model offers a natural explanation for the bumps and fluctuations in light curves (e.g., \citealt{Inserra2017, handbook, Moriya2018}), but the absence of relatively narrow emission lines in the spectra of most SLSNe-I is often regarded as a major issue for the CSM model.

Gaia17biu/SN~2017egm (with a redshift $z=0.03063$) was discovered by the Photometric Science Alerts Team of the {\it Gaia} mission on 2017 May 23 (UT dates are used throughout this paper) and subsequently classified as an SLSN-I (\citealt{Dong2017}), which was the closest SLSN-I discovered by then. In contrast to the typical dwarf and metal-poor host galaxies for SLSNe-I, its host galaxy, NGC~3191, is massive \citep{Dong2017} and has a mean metallicity near solar (\citealt{Nicholl2017a}, \citealt{Bose2018}; see also spatially resolved observations by \citealt{Chen2017}, \citealt{Izzo2018}). Its pre-peak near-ultraviolet (NUV) to optical colors \citep{Bose2018} are similar to those of Gaia16apd \citep{Kangas2017} and among the bluest observed for an SLSN-I, while its peak luminosity ($\rm M_g = -21$\,mag) is substantially lower than that of Gaia16apd. Moreover, SN~2017egm holds the tightest upper limit of radio luminosity among SLSNe-I \citep{Bose2018}. The sharp peak in the light curves of SN~2017egm are shown to be well fitted by ejecta-CSM interaction (CSI) models, but are difficult to explain by only a magnetar or radioactive decay of $\rm ^{56}Ni$ (\citealt{Wheeler2017}). \citet{Hosseinzadeh2022} also concluded that SN~2017egm cannot be explained well by the magnetar model alone, given the multiple bumps in its extended post-peak light curve.

As one of the closest SLSNe-I and having extensive multiwavelength observations, SN~2017egm appears to be an ideal target for further detailed studies, which may help us understand the explosion mechanism of SLSNe-I. In this paper, we present the most complete photometric and spectroscopic analysis of SN~2017egm, taking our observational data and all public data together.
We present the observations and data reduction in Section 2. In Section 3, we analyze its light curves and spectra, and we classify it as an SLSN-Ib and compare it to other SLSNe-I. Section 4 presents a discussion and our conclusions. In this work, we adopt a luminosity distance of $D_L=138.7\pm1.9$\,Mpc assuming a standard {\it Planck} cosmology (\citealt{Planck2016}).

\begin{figure*}[htb]
\figurenum{1}
\centering
\begin{minipage}{1.0\textwidth}
\centering{\includegraphics[angle=0,width=0.8\textwidth]{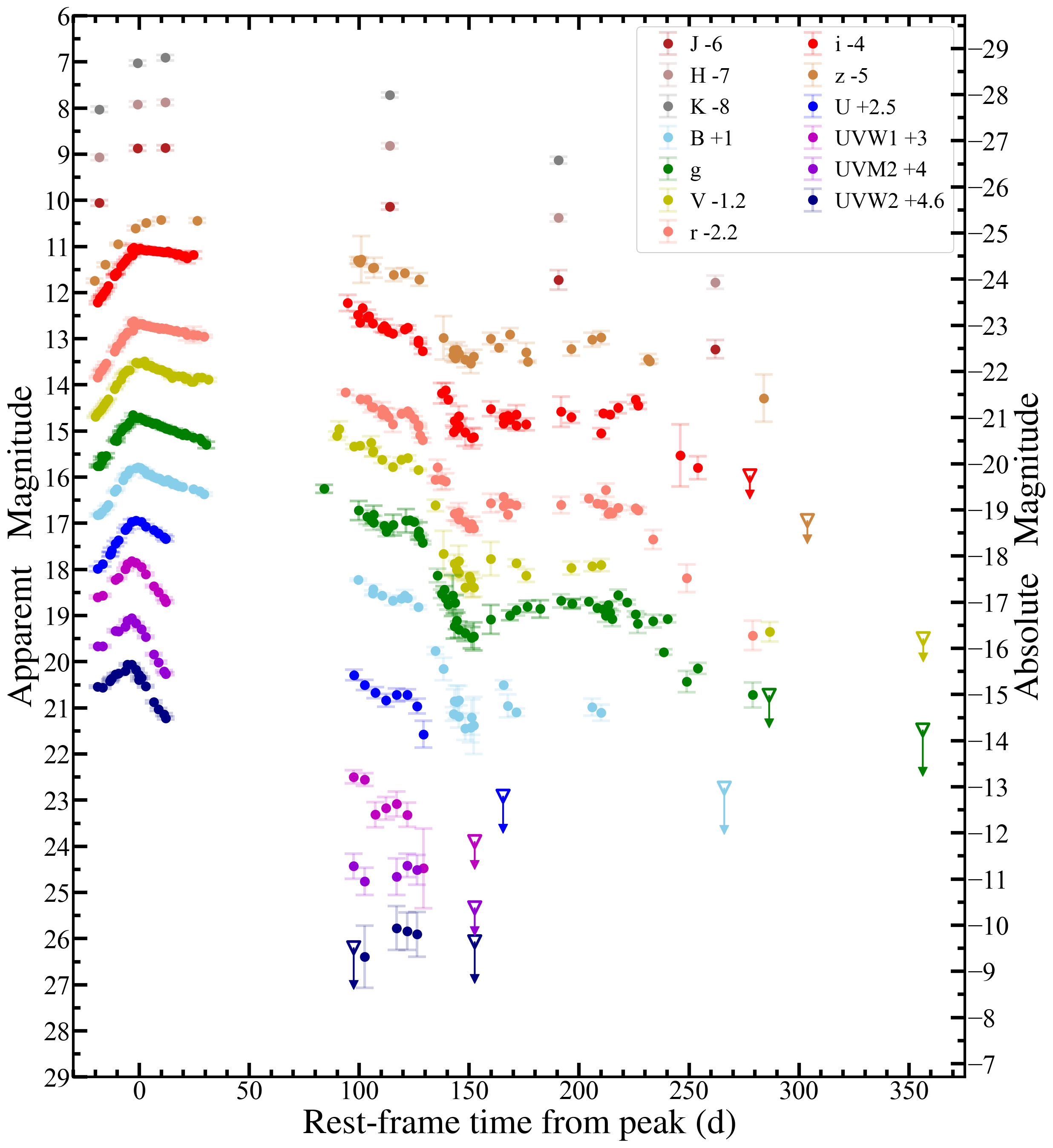}}
\end{minipage}
\caption{Multi-band light curves of SN~2017egm from NUV to NIR with Galactic extinction corrected. The light curves are shifted vertically at various offsets for different bands for display. The reference epoch is set to be the $g$-band peak at JD = 2,457,926.3. All optical light curves show a rapid descent during 120$-$150 days and multiple bumps distinctly. The hollow triangles with downward arrows represent upper limits of the corresponding bands. The data used to create this figure are available in machine-readable form.\label{LCs}}
\end{figure*}

\section{Observations and Data}

Multiband photometry from UV to mid-infrared (MIR) wavelengths, as well as optical spectra spanning $\sim 300$ days since the discovery of SN~2017egm, were taken by us or gathered from publicly-available archives. Here we describe the relevant observations and data reduction. In addition, analysis of the four epochs of \emph{Chandra} visits are presented for the first time.

\subsection{Ground-Based Optical and Near-Infrared Photometry}

We obtained late-time multiband images with a set of ground-based instruments. Optical data in the $BVgriz$ bands were taken with 0.6\,m telescopes at Post Observatory (PO) at the Sierra Remote Observatories (SRO; CA, USA), the 2.0\,m Las Cumbres Observatory Global Telescope network (LCOGT), the 2.0\,m Liverpool Telescope (LT) at La Palma, and the ALFOSC mounted on the 2.6\,m Nordic Optical Telescope (NOT). The $JHK$ near-infrared (NIR) images were obtained with the NIR Camera mounted on the NOT (NOTCam) and the NIR Wide-Field Camera mounted on the United Kingdom Infrared Telescope (UKIRT).  

We found that the transient signals associated with SN~2017egm were undetectable after $\sim360$\,days in all of our optical and NIR images. Thus, we generate the reference images with these late-stage data, which are dominated by host-galaxy emission around the position of SN~2017egm. All photometric measurements were performed on the difference images, which are obtained by subtracting the reference image from a current science image by matching the point-spread function (PSF). Specifically, we used {\tt HOTPANTS}\,\citep{Becker2015} for image subtraction. Before subtraction, we removed cosmic rays and aligned the images using Astrometry.net. Most instruments have good references with high signal-to-noise ratios (SNRs) and Gaussian-like PSF profiles, so the subtractions were conducted directly with their corresponding references. However, the quality of the final-epoch LT image did not satisfy the requirement for a reference image owing to clouds; thus, the 2.0\,m LCOGT reference images were adopted instead.

After image subtraction, PSF photometry was performed on the difference image with the Photutils package of Astropy \citep{Astropy} for the optical and NIR data. The photometric data were calibrated using PS1 standards \citep{PS1} in the field of view (FOV) with Pan-STARRS color-term corrections \citep{Tonry2012} applied for the Johnson $BV$ filters and SDSS $griz$ filters, and the $JHK$ NIR data were calibrated using 2MASS standards.  

We reprocessed early-time optical data observed by the above-mentioned instruments presented by \citet{Bose2018} for consistency according to the same procedure in this work. The phases in the manuscript are given in the rest frame of SN~2017egm, and the reference epoch is set to be the $g$-band peak at JD = 2,457,926.3.

\subsection{Swift/UVOT photometry}

UV images were obtained with the {\it Neil Gehrels Swift Observatory} (hereafter {\it Swift}) with the Ultra-Violet Optical Telescope (UVOT). The {\it Swift}/UVOT data covered $\sim 140$ days of NUV data after discovery, which provided good sampling for a blue bump before the NUV became undetectable (see Figure \ref{LCs}). The {\it Swift}/UVOT photometry was measured with UVOTSOURCE task in the {\tt Heasoft} package using 5\arcsec apertures after subtracting the galaxy background using a {\it Swift}/UVOT image taken on 2018 June 26 (+360 days). It was placed in the AB magnitude system \citep{OkeGunn83}, adopting the revised zero points and sensitivity from \citet{Breeveld2011} as done in the previous work analyzing the early-time UVOT data (\citealt{Bose2018}). All of the photometric data are presented in Table \ref{photometry_table}.

\subsection{Archival MIR Data}

The \emph{Wide-field Infrared Survey Explorer} (\wise) conducted a full-sky survey at 3.4, 4.6, 12, and 22\,$\mu$m (labeled as W1--W4, respectively) from 2010 February to August (\citealt{Wright2010}). The solid hydrogen cryogen used to cool the W3 and W4 instruments was depleted later and placed in hibernation in 2011 February.
\wise\ was reactivated and renamed \neowise-R in late 2013, using only W1 and W2, to search for asteroids that could pose potential impact hazards to Earth (\citealt{Mainzer2014}).
Up to 2021 December, the \neowise\ survey has visited every region of the sky 16 times every six months, with an average of 12 single exposures within each epoch (typically within one day).

Previous works show that the IR echoes (UV/optical light absorbed and re-radiated in the IR by dust) of optical transients are detectable on time scales of months to years while the intraday variability is negligible (e.g., \citealt{Jiang2016, Jiang2021,Sun2022}). 
To acquire high-SNR measurements, we choose to perform photometry on the time-resolved \wise/\neowise\ coadds, and each coadd per band per epoch is produced by stacking exposures within 
typically $\sim 1$\,day intervals (\citealt{Meisner2018})\footnote{\url{https://portal.nersc.gov/project/cosmo/temp/ameisner/neo8}}. 

\begin{figure}
\figurenum{2}
\centering
\begin{minipage}{0.5\textwidth}
\centering{\includegraphics[width=1\textwidth]{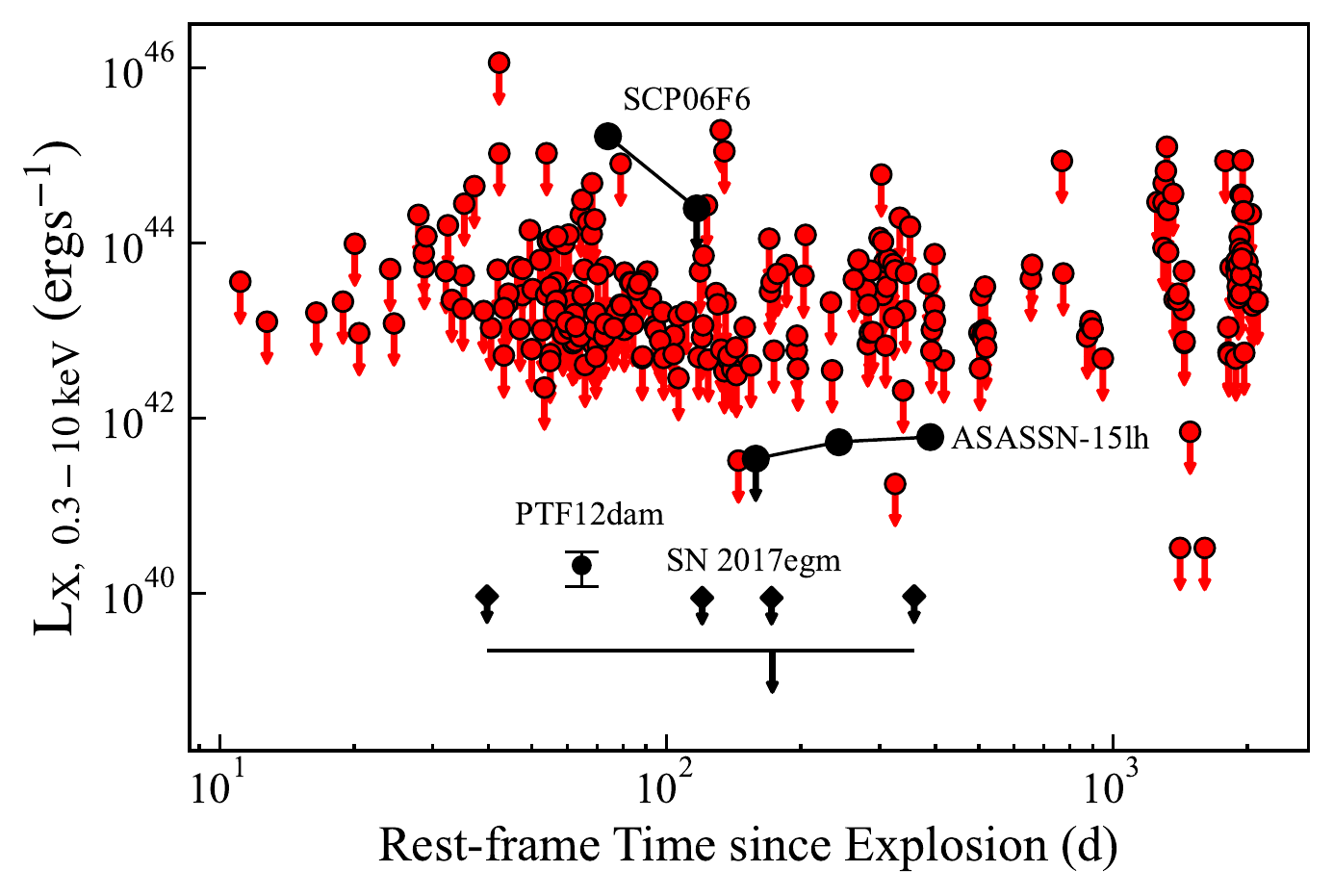}}
\end{minipage}
\caption{X-ray observations of SN~2017egm along with other SLSNe-I spanning the phase range $\sim10$--2000\,d (red circles for upper limits, black circles for detections or possible detections from \citealt{Margutti2018}; black diamonds for SN~2017egm) since explosion. Black horizontal line with arrows is the SN~2017egm average weighted by exposure time. The explosion time of SN~2017egm is estimated as the time of its first detection. \label{x}}
\end{figure}

\subsection{X-Ray Observations}

Four epochs of X-ray observations of SN~2017egm (PI R. Margutti) were obtained with the \emph{Chandra X-ray Observatory (CXO)} on 2017 June 26 (24\,ks exposure), 2017 September 17 (25\,ks), 2017 November 9 (25\,ks), and 2018 May 21 (24\,ks); they correspond to +4, +85, +136, and +323 days since the optical peak. We processed the {\it Chandra} data with CIAO (v.4.14) using the latest calibration files (v.4.9.6). The Level-2 event files were recreated by the script of {\tt chandra\_repro}. SN~2017egm is not detectable at any individual epoch. Then we analyzed the stacked images of all four epochs with a total exposure time of 96.9~ks, still yielding a non-detection. Assuming a power-law spectrum with photon index $\Gamma=2$ and Galactic column density $N_{\rm H}=1.08\times10^{20}\,\rm cm^{-2}$, the inferred 3$\sigma$ unabsorbed flux upper limit is $8.4\times10^{-16}$\,\flux\ in the 0.5--10\,keV band, corresponding to $1.9\times 10^{39}$\,\lum\ at the distance of SN~2017egm. Alternatively, if we assume a thermal emission, e.g., a blackbody with temperature of $kT=1$\,keV, the inferred unabsorbed flux upper limit is $7\times10^{-16}$\,\flux, corresponding to a luminosity of $1.8\times10^{39}$\,\lum. 

The non-detection from the deep \emph{Chandra} observations gives the most stringent constraint on the X-ray emission of an SLSN-I to date. Figure~\ref{x} shows this limit compared with a large sample of X-ray observations of other SLSNe-I from \citet{Margutti2018}. The X-ray of SN~2017egm could be heavily absorbed by CSM, i.e., a neutral hydrogen column density of $\nh=10^{24}\,\rm cm^{-2}$ will lead to a non-detection of intrinsic X-ray luminosity of $10^{41}$\,\lum.

\subsection{Spectroscopic Observations}

Optical spectra from $+84$\,d to $+261$\,d were obtained using the Kast Spectrograph mounted on the 3\,m Shane telescope at Lick Observatory (CA, USA; Miller \& Stone 1993), the Double Spectrograph for the Palomar 200-inch Hale telescope (P200), the MODS1 multi-object double spectrographs mounted at the Large Binocular Telescope Observatory (LBT), and the OSIRIS instrument located in the Gran Telescopio CANARIAS (GTC). We also obtained spectra of SN~2017egm in the NIR at +105 days using the Aerospace Corporation's Visible and Near-Infrared Imaging Spectrograph (VNIRIS) mounted on the Shane telescope and +143 days using the Spex medium-resolution spectrograph (0.7--5.3~$\mu$m; \citealt{Rayner2003}) on the NASA Infrared Telescope Facility (IRTF). All of the spectra except for the one observed with LBT/MODS were taken at or near the parallactic angle to minimize differential slit losses caused by atmospheric dispersion (\citealt{Filippenko1982}).
Regarding the LBT observation, the slit width was 1\,arcsec and the slit position angle was set to the default 0 deg (north-south), which corresponds to a parallactic angle of 146 deg at the start of observation. However, due to a low airmass ($\sim$ 1.0) and short exposure time (600\,secs), the effect of differential atmospheric refraction should be small (see \citealt{Filippenko1982}).
The LBT/MODS data were processed with the modsCCDRed \citep{2019zndo...2550741P} program and the MODSIDL\citep{2019zndo...2561424C} pipeline. The raw measurements of SpeX data were converted to absolute flux using the Spextools software package (\citealt{Vacca2003,Cushing2004}). The VNIRIS data were processed using software customized for VNIRIS and written in the Interactive Data Language (\citealt{Rudy2021}). All of the other spectra were reduced and calibrated using standard procedures in IRAF\footnote{IRAF is distributed by the National Optical Astronomy
Observatories, which is operated by the Association of Universities for
Research in Astronomy, Inc. (AURA) under cooperative agreement with the National
Science Foundation.}. See a summary of our late-time spectra in Table~\ref{spec_table}.

\section{Analysis and Results}

\subsection{Photometric and Color Evolution}

All light curves are shown in Figure \ref{LCs}, including our data presented in this work along with early-phase data from \citet{Bose2018} and late-phase $gri$ data published by \citet{Hosseinzadeh2022}. The magnitudes are corrected for Galactic extinction ($R_V=3.1$ and $E(B-V)=0.0097\pm0.0005$\,mag; \citealt{Schlafly2011}) following \citet{Bose2018} and $K$-corrections based on the optical spectra.

During $\sim2$ weeks following the season gap ($\sim+30$--100\,d), the light curves are in a relatively smooth decline, which appears to extend the post-peak linear decline in magnitudes prior to the gap. Its post-peak decline rate is one of the slowest among SLSNe-I \citep{Bose2018}, but its peak luminosity ($M_{g,{\rm peak}} \approx -21$\,mag) is close to the mean for SLSNe-I, defying the empirical correlation reported by \citet{Inserra2014} that the more slowly declining SLSNe-I tend to be more luminous.

At $\sim+115$\,d, the optical light curves show a small bump in all bands, followed by a rapid decline ($\sim 2$ mag) at $\sim +130$--150\,d. Meanwhile, the $U$-band light curve also exhibits a bump followed by possibly an even more rapid drop of NUV fluxes given the non-detection around +150\,d. The bump is reminiscent of that found in SN~2015bn (\citealt{Nicholl2016}) which was more pronounced in the bluer bands. A sharp decline in all bands a few months after the peak is an expected common feature for the ejecta-CSM interaction model, and this occurs when the shock wave reaches the outer edge of the extended envelope with no more material ahead, with no source of radiation energy remaining \citep{Sorokina2016}. 
Interestingly, there is a second bump $\sim+ 150$--240\,d, which is most clearly seen in the $g$-band light curve. The second bump appears nearly flat, and the variations are particularly small in redder bands, suggesting an extra power supply, such as CSM interaction.

\begin{figure}
\figurenum{3}
\flushleft
\begin{minipage}{0.5\textwidth}
\flushleft{\includegraphics[angle=0,width=1\textwidth]{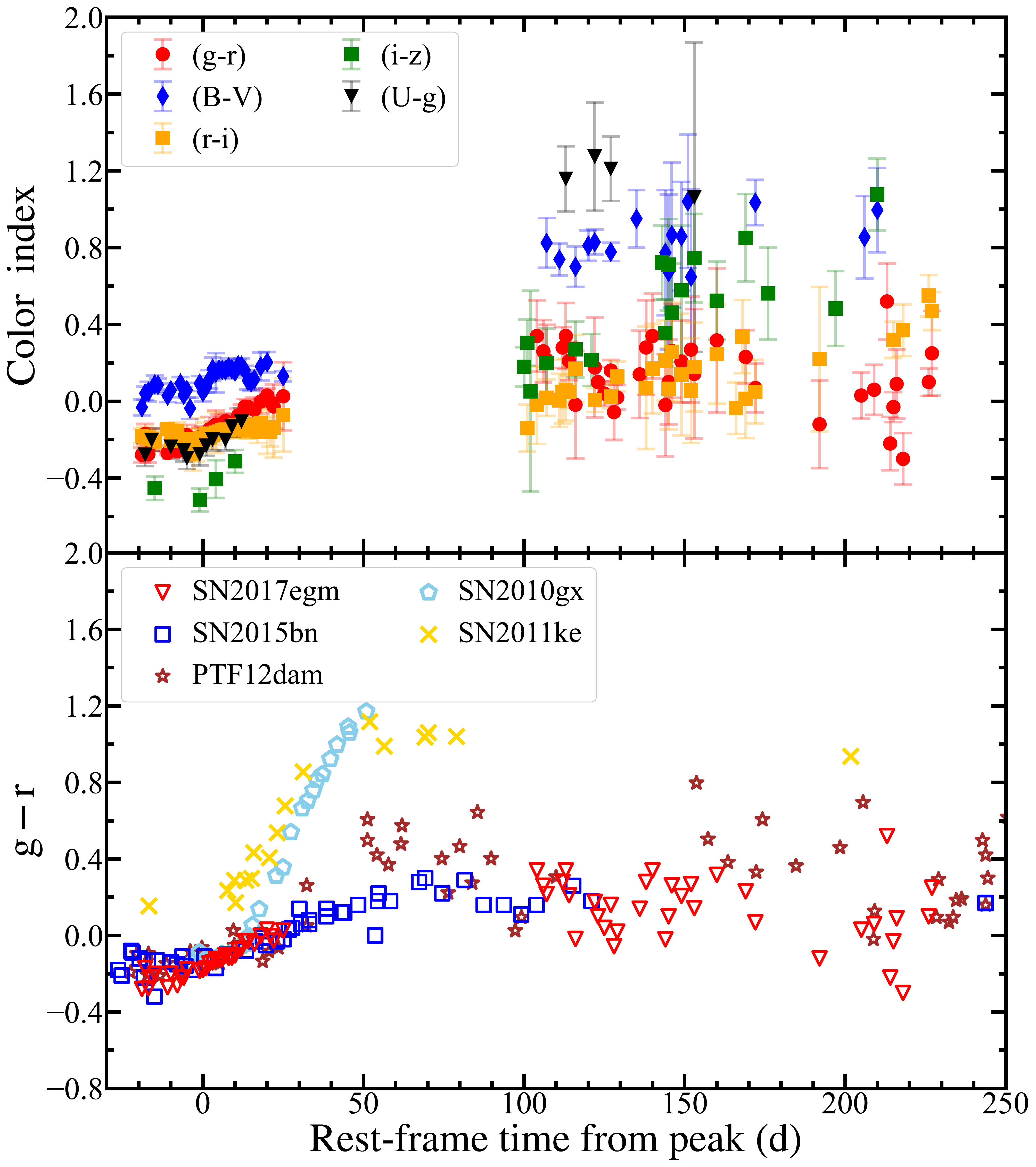}}
\end{minipage}
\caption{Optical color evolution of SN~2017egm.
{\it Top:} evolution of a selection of optical colors.. 
{\it Bottom:} comparison of rest-frame $g-r$ color with that of other well-observed SLSNe-I.\label{color}}
\end{figure}

As shown by \citet{Bose2018}, the UV-to-optical color ($uvw2-r$) evolution near the peak of SN~2017egm is similar to that of Gaia16apd  (\citealt{Kangas2017}), which starts with a very blue color and becomes redder rapidly as the SN cools. By contrast, the optical color evolution is much less dramatic  (see the top panel of Figure~\ref{color}). We compare the $g-r$ color evolution of SN~2017egm with that of other well-observed SLSNe-I and find it is most similar to SN~2015bn (see the bottom panel of Figure~\ref{color}) during all available phases. They both show less evolution in comparison with others. Their optical colors are roughly constant near the peak with $g-r \approx r-i \approx -0.2$\,mag, and they become gradually redder until $\sim+100$\,d.  After $\sim+100$\,d, the $g-r$ color curve of SN~2017egm appears to flatten out, probably because of multiple bumps and the strong, varying emission-line features in the corresponding wavelength regimes.

\subsection{Pseudo-bolometric Light Curve}

The densely-sampled multiband light curves of SN~2017egm allow us to analyze its spectral energy distributions (SEDs) at different epochs. We use {\tt PhotoFit} package (\citealt{Soumagnac2020}) to fit the blackbody temperature ($T_{\rm bb}$), luminosity ($L_{\rm bb}$), and effective radius ($R_{\rm bb}$). The fitting results are given in Table~\ref{bbfit_table}. The blackbody model fits well the SEDs near maximum light. The observing cadence after +150 days in $B$ and $V$ is sparse except for the $g$ band; thus, we interpolate the $g$-band light curve to match with the $BV$ epochs. We require at least three photometry bands in the fitting. We do not use the NIR $JHK$ data in the single blackbody fitting, and their analysis is shown in next section.

\begin{figure}
\figurenum{4}
\centering
\begin{minipage}{0.5\textwidth}
\centering{\includegraphics[width=1\textwidth]{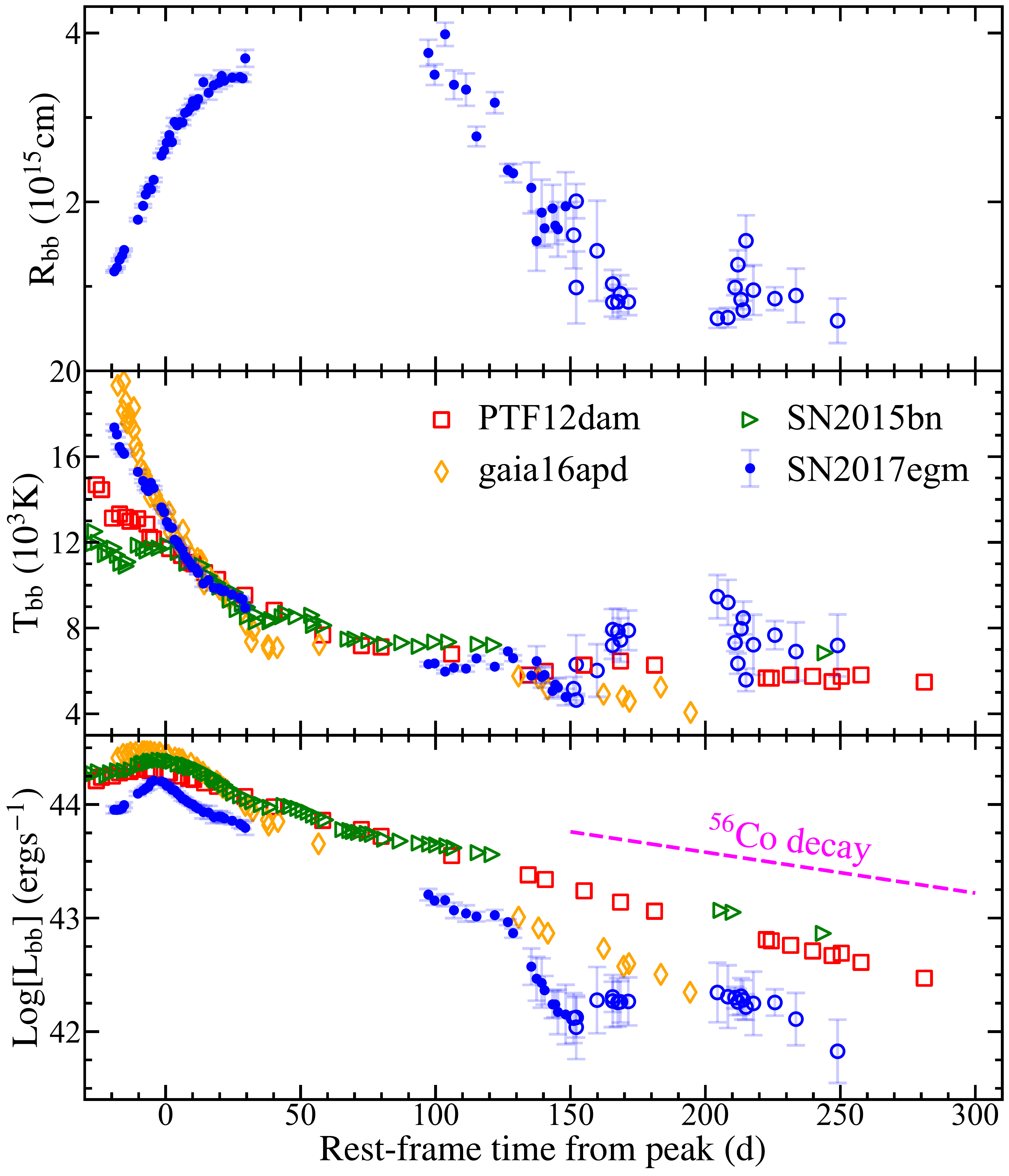}}
\end{minipage}
\caption{Evolution of the blackbody radius {\it (top panel)}, temperature {\it (middle panel)}, and luminosity {\it (bottom panel)} of SN~2017egm. The evolution of three other well-observed SLSNe-I [PTF12dam (red squares), SN2015bn (green triangles), and Gaia16apd (orange diamonds)] is overplotted for comparison.
The open symbols denote that the blackbody fitting is performed without NUV photometry.\label{bbfit}}
\end{figure}

In Figure~\ref{bbfit}, we show the temporal variations of inferred blackbody parameters and compare them with those of PTF12dam (\citealt{Nicholl2013}), Gaia16apd (\citealt{Kangas2017}), and SN~2015bn (\citealt{Nicholl2016}). SN~2017egm evolves in temperature like Gaia16apd but with smaller radii and luminosities. The total observed radiation energy is close to $7.8\times10^{50}$\,erg. Distinct from the smooth decline in the other three, the bolometric light curve of SN~2017egm features several slope changes, including high temperature initially with a steady cooling, a small bump followed by a steep decline, and then a late-time broad bump accompanied by rising temperature. It is worth noting that the late-time bolometric light curves at $\gtrsim 150$\,d, which exhibited variations including a broad bump, were synthesized without NUV photometry. Thus, the derived blackbody parameters were less certain than those at earlier phases.  However, such a multitude of variations with complicated characteristics suggests that it may have undergone multiple physical processes, which can hardly be explained by a simple magnetar or $\rm ^{56}Ni$ decay model. 

Recent work suggests that the undulations and late-stage bumps appear to be a common phenomenon among SLSNe-I (\citealt{Inserra2017,Chen2022,Hosseinzadeh2022}). There are two main interpretations. One attributes them to the temporal variation of the central engine, which could be modulated by photon diffusion in the ejecta, or the change in opacity that causes the variations in photospheric emission (see discussions in \citealt{Hosseinzadeh2022} and references therein). The other mechanism invokes the CSI model, in which the ejecta-CSM interaction is an effective process to convert mechanical energy into thermal emission. In the case of SN~2017egm, the latter scenario seems to be preferable. First, \citet{Wheeler2017} find that the linear rise and decline around the sharp peak of SN~2017egm can be well explained by the CSI model, but is difficult to be fitted by the magnetar model, and they also expected that the late-time light curves could discriminate between the two models. The magnetar model \citep{Vurm2021, Hosseinzadeh2022} has significant issues to fit the bumps. We make further discussions on the models in Section 4. 

\begin{figure}[htb]
\figurenum{5}
\centering
\begin{minipage}{0.5\textwidth}
\centering{\includegraphics[width=1\textwidth]{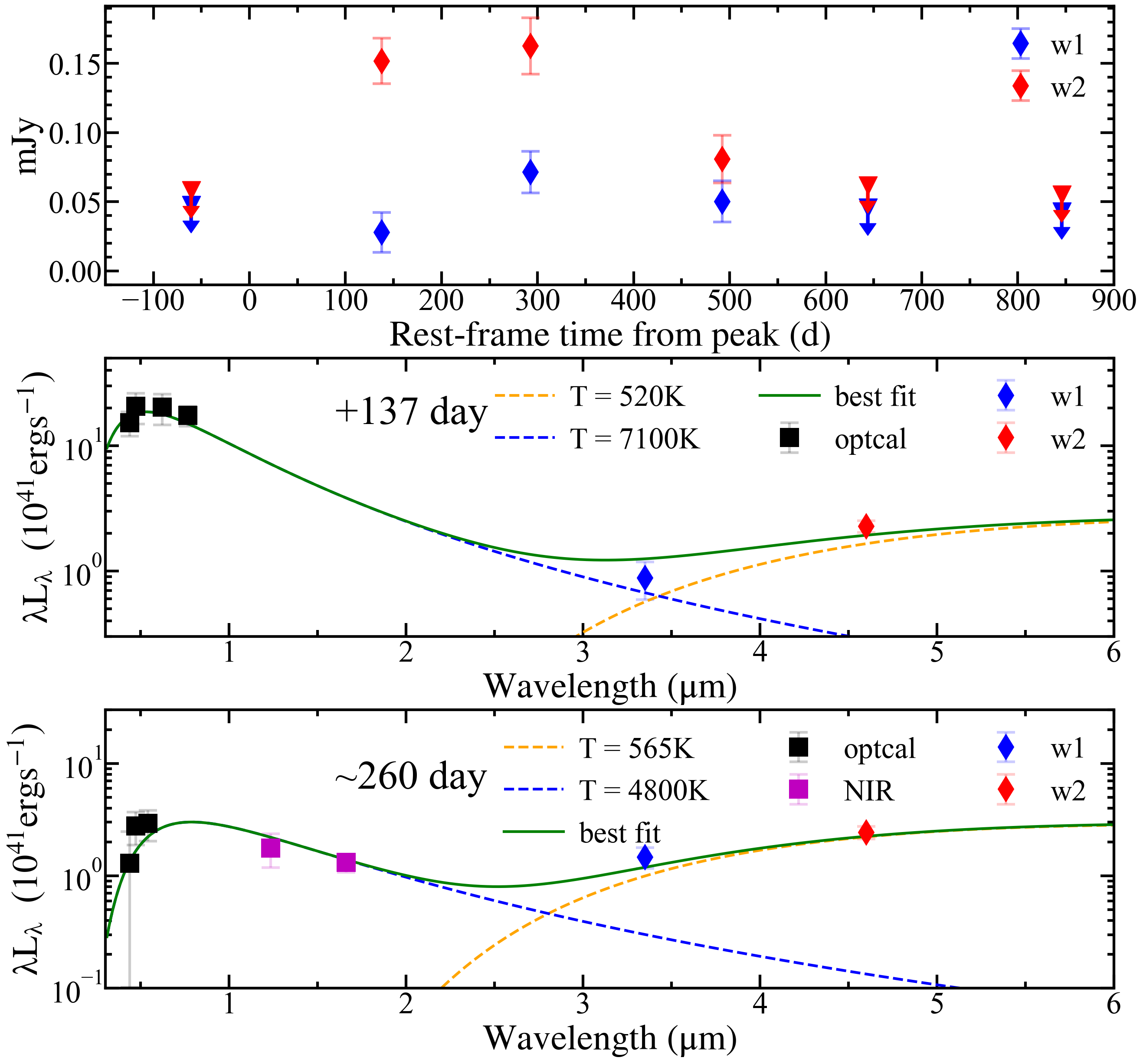}}
\end{minipage}
\caption{MIR light curves of SN~2017egm and the double-blackbody fitting of the UV-optical-IR SEDs.
{\it Top panel:} W1 (3.4\,$\mu$m, blue diamonds) and W2 (4.6\,$\mu$m, red diamonds) light curves of SN~2017egm measured from the NEOWISE survey, in which the triangles mark the $3\sigma$ upper limits.
{\it Middle and bottom panels:} SED fitting at +137 and +260 days. Two blackbody components are required to account for the overall SED, which is peaked at optical and MIR bands, respectively.
\label{mir}}
\end{figure}

\begin{figure*}[ht]
\figurenum{6}
\centering
\begin{minipage}{1\textwidth}
\centering{\includegraphics[width=0.95\textwidth]{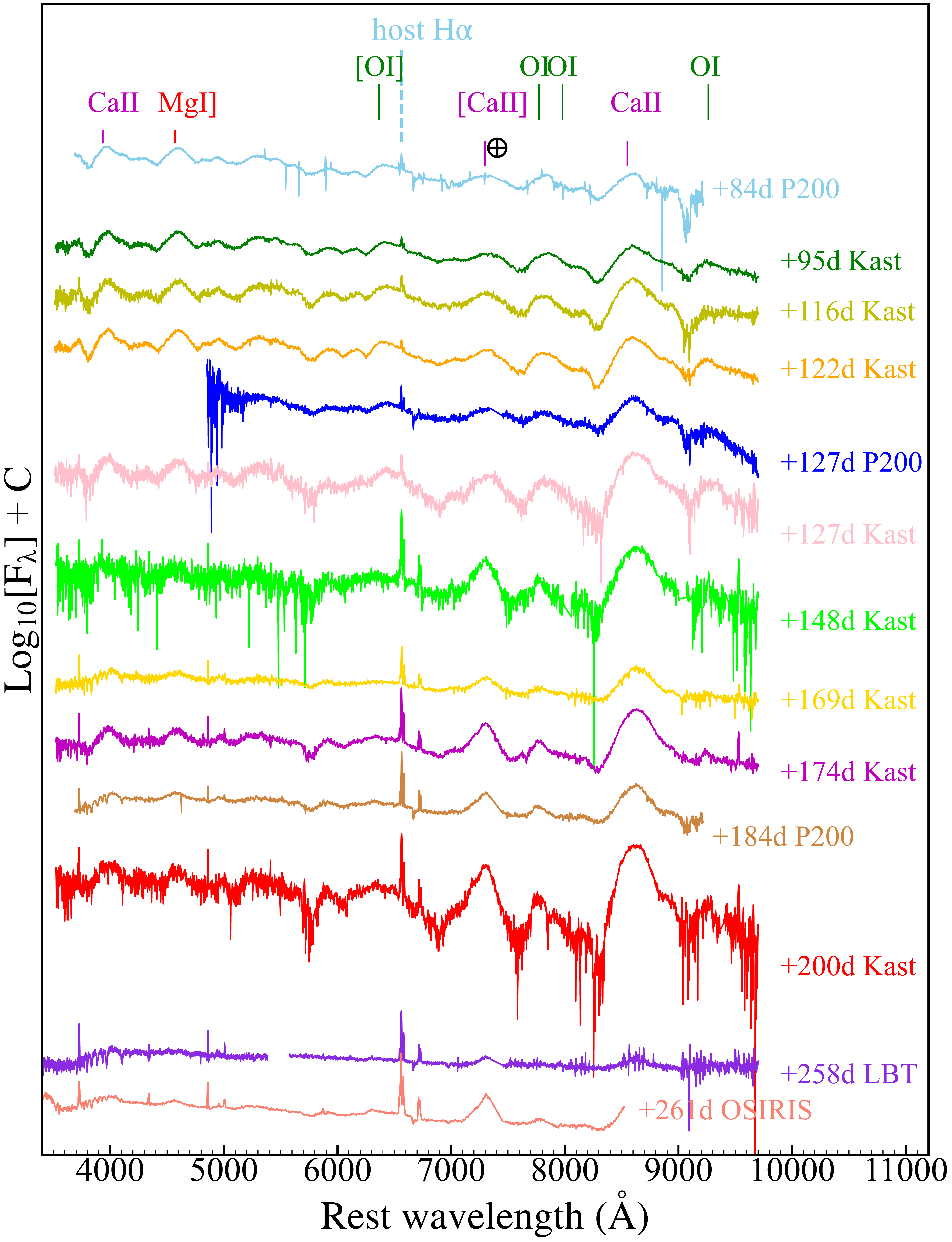}}
\end{minipage}
\caption{Rest-frame spectral evolution of SN~2017egm. Strong O~I , Ca~II, and Mg~I] $\lambda$4571 features are marked by different colors at the top. The $\oplus$ symbol marks the positions of the strongest telluric absorption bands which have been masked. Each spectrum is labeled by the instrument employed and the rest-frame phase from maximum light. The data used to create this figure are made available digitally. \label{spec_all}}
\end{figure*}

\subsection{MIR Excess and Dust Emission}

The NEOWISE survey visited the sky region of SN~2017egm every six months since 2014 and has acquired a total of 14 epochs of observations. The images at the first NEOWISE epoch are used as references since they were observed securely before the SN explosion. We then perform image subtraction with {\tt HOTPANTS} and PSF photometry on the difference images following \citet{Jiang2021}. We take the epochs displaying $>3\sigma$ detection at the position of SN~2017egm in at least one band as robust MIR excess (see light curve in Figure~\ref{mir}). The excesses are clearly detectable at +137, +292, and +492 days after the $g$-band peak. SN~2017egm remained luminous in the optical band at +137 and +260 day, indicating a non-negligible contribution from the Rayleigh-Jeans tail of the optical blackbody. Therefore, we try to fit the overall optical-NIR-MIR SEDs with double-blackbody functions for these two epochs (see Figure~\ref{mir}). The fitting results show that the MIR excesses are mainly dominated by the blackbody component peaked in the MIR, which can be naturally attributed to reprocessed dust emission. The dust blackbody temperatures at the three epochs are $520\pm70, 565\pm83$, and $827\pm45$\,K with luminosities of $(3.54\pm1.50)$, $(3.89\pm2.01)$, and $(1.65\pm0.40) \times 10^{41}$\,erg\,s$^{-1}$, respectively (see Table~\ref{IR_bb_table} for the fitting results). The last epoch has a significantly bluer W1-W2 color and subsequently higher inferred blackbody temperature compared to the previous two epochs, suggesting that the dust might be heated up while the SN luminosity decreased.
Our measurements of the W1 and W2 excesses are  consistent with those found by \citet{Sun2022}, but they did not infer blackbody parameters of dust emission owing to a lack of contemporaneous optical data.

The origin of dust in SLSNe remains poorly explored until very recently. \citet{Chen2021} has concluded that the prominent IR excess detected after +230d in SN~2018bsz can be only well explained by newly-formed dust within the SNe ejecta after careful analysis. Moreover, \citet{Sun2022} performed a systematic study of the MIR light curves of 10 SLSNe at $z<0.12$, including SN~2017egm, and suggested that the MIR excess of SN~2017egm may also be explained by the newly-formed dust model as SN~2018bsz owing to their similarities. One notable characteristic that makes SN~2017egm distinct from other slowly declining SLSNe-I is the rapid luminosity drop that occurred between +120\,d and +150\,d, which is reminiscent of the sudden drop starting at +98\,d observed in SN~2018bsz. However, as claimed by \citet{Chen2021}, the drop is yet likely due to the ejecta transitioning to the nebular phase rather than dust extinction although robust dust formation evidence appears later. Actually, an IR echo of a proper pre-existing spherical dust shell can equally reproduce the observed MIR emission. Note that $R_{\rm bb, IR}$ is around (7.3-8.2)$\times10^{16}$cm, which is much larger than $R_{\rm bb, Opt}$ ($\sim10^{15}$cm) in our double-blackbody fits at both epochs. The IR emission comes much further out than the region of ejecta, and thus the excess may be dominated by pre-existing dust emission. It is worthwhile to note that the inferred dust temperature shows a rising trend albeit with large errors, that is unusual in the echo scenario. Nevertheless, an asymmetric dust bubble shell, e.g., shell in the far side is closer to the SNe than that in our line of sight, can produce such a trend. In spite of this, we can not definitively exclude a  dust formation process in later epochs, particularly considering a higher blackbody temperature and smaller radius for the third MIR epoch (+492\,d, see Table~\ref{IR_bb_table}), which may represent the dust formed during ejecta-CSM collision. In fact, the newly-formed and pre-existing dust could work simultaneously, as has been demonstrated in SN~2006jc (\citealt{Mattila2008}) and some other SNe~II (e.g., \citealt{Pozzo2004, Meikle2007}).

\subsection{Spectroscopic Analysis}

The optical spectra are presented chronologically in Figure~\ref{spec_all}, with notable O~I, Ca~II, and Mg~I] emission lines marked.  For $\sim 200$ days after the peak, its blue continuum faded rapidly, and in contrast, the emission-line features evolve slowly. We make the line identifications by comparing with other SLSNe-I (\citealt{Nicholl2016}; \citealt{Kangas2017}; \citealt{Quimby2018}; \citealt{Nicholl2019}).

\begin{figure}
\figurenum{7}
\centering
\begin{minipage}{0.5\textwidth}
\centering{\includegraphics[width=1\textwidth]{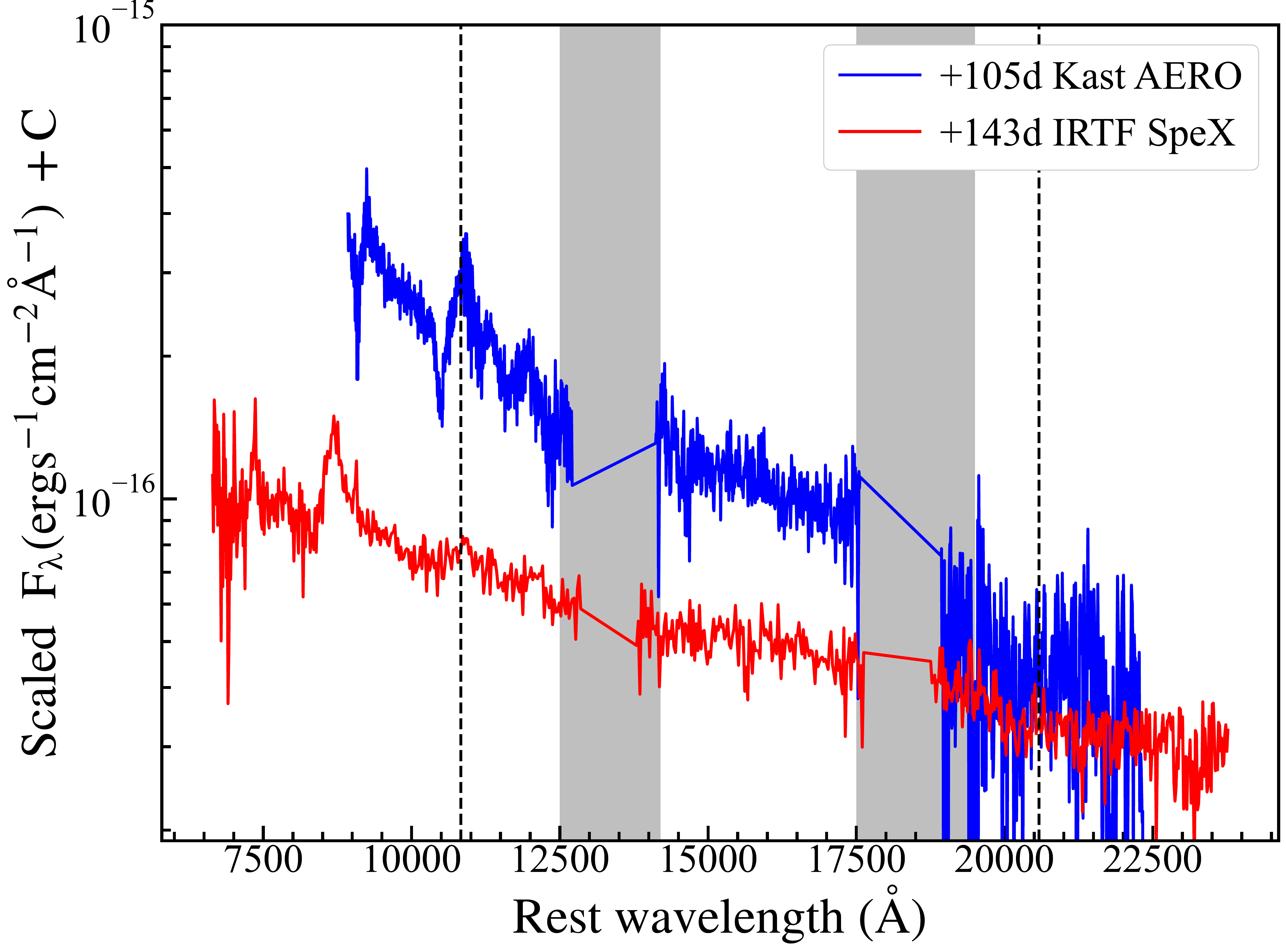}}
\end{minipage}
\caption{NIR spectra at two epochs. A prominent He~I line $\lambda$10,830 is detected at +105\,d. Strong
telluric and unreliable regions in the spectra are masked out. 
The dashed lines mark the rest-frame wavelengths of He~I lines at 10,830\,\AA\ and 20,586\,\AA.
The data used to create this figure are made available digitally. \label{IRspec}}
\end{figure}

Two NIR spectra are presented in Figure~\ref{IRspec}. There is a strong feature around the region of He~I $\lambda$10,830 at +105\,d. There was controversy in the past for the identification of He~I $\lambda$10,830 in SLSN-I spectra. The proposed He~I $\lambda$10,830 in the spectra of SN~2012il \citep{Inserra2013} was later found to be a broad emission feature redshifted by $\sim1500$\,km\,s$^{-1}$ relative to $\lambda$10,830, and may instead be blueshifted nebular Pa$\gamma$ emission \citep{Quimby2018}. In our case, the feature around 10,830\,\AA\ at +105\,d also shows a strong P~Cygni profile with a velocity of $\sim -8000$\,km\,s$^{-1}$, that are similar with SN~2019hge, the only other SLSN-I with HeI 1.08um, so it cannot be a nebular Pa$\gamma$ line. The non-detection of He~I at $\sim 2.05\,\mu$m might be simply due to the low SNR, since it is generally much weaker than He~I $\lambda$10,830 \citep{Yan2020}. Interestingly, the He~I feature faded rapidly with the detection of a weak emission line at +143 days. On the other hand, it is worthwhile to note that there is no significant He~I $\lambda$10,830 in the early-time NIR spectra ($-2.7$ and $-0.5$ days) according to \citet{Bose2018}, indicating that the He~I could be short-lived. 

\begin{figure*}[htb]
\figurenum{8}
\centering
\begin{minipage}{1\textwidth}
\centering{\includegraphics[width=1\textwidth]{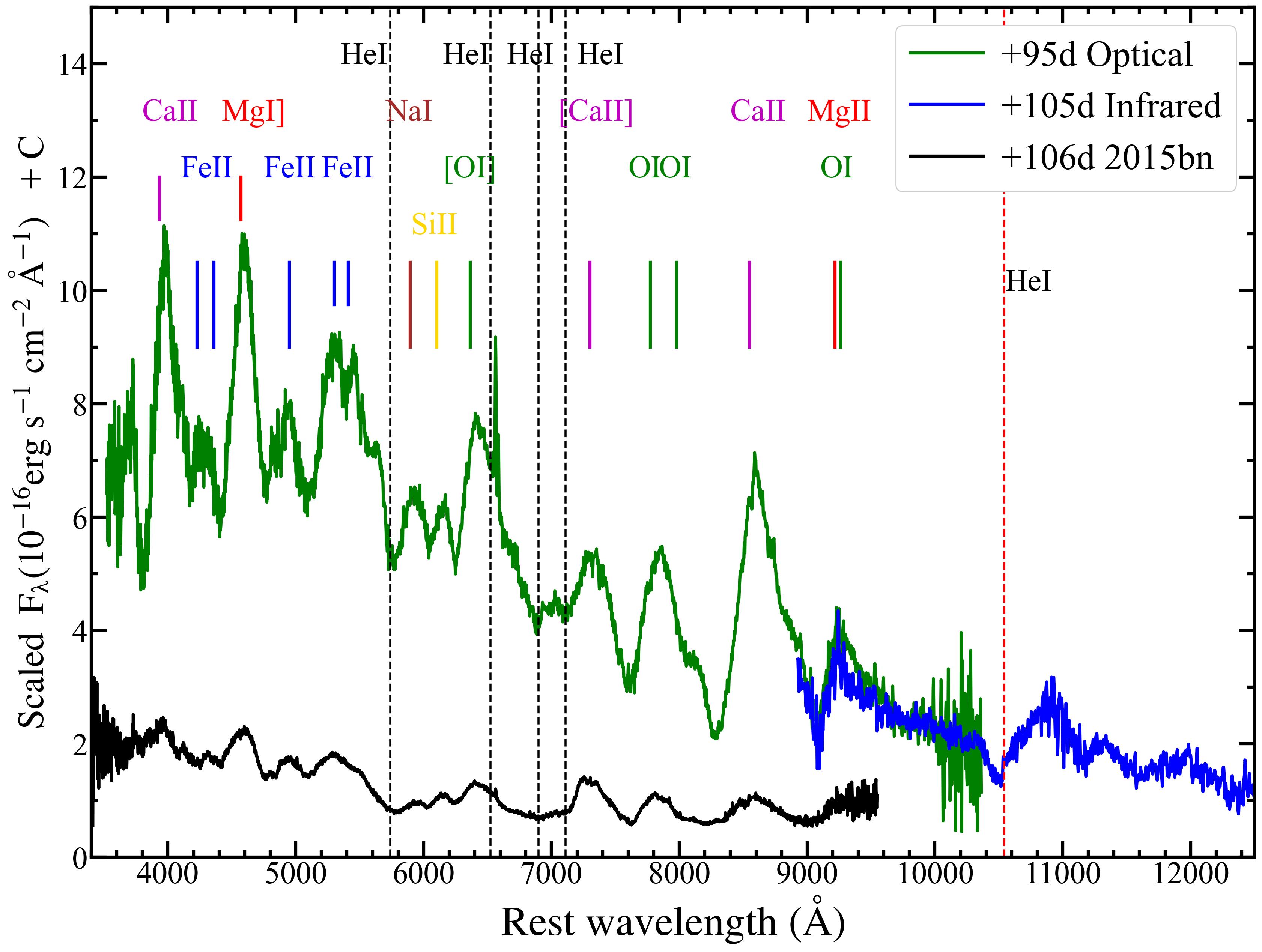}}
\end{minipage}
\caption{Optical spectrum of SN~2017egm at +95 days post-maximum and NIR spectrum at +105 days. The optical spectrum of SLSN-Ic SN~2015bn at +106 day is plotted for comparison. The identified lines are labeled by different colors at the top. The dashed lines mark the locations of He~I $\lambda\lambda$5876, 6678, 7065, and 7281, with a velocity of $-7000$\,km\,s$^{-1}$ (black) and He~I $\lambda$10,830 with a velocity of $-8000$\,km\,s$^{-1}$ (red). \label{line}}
\end{figure*}

\citet{Yan2020} reported six He-rich SLSNe-Ib from the ZTF Phase-I SLSN-I sample, suggesting that the key to identifying He-rich events is the presence of multiple He features. We cross-check our optical spectra for He features. In Figure~\ref{line}, we compare the spectra of SN~2017egm at $\sim +100$\,d with the well-studied SLSN-Ic SN 2015bn in detail (see Figure~\ref{line}). Four He~I absorption lines (He~I $\lambda\lambda$5876, 6678, 7065, 7281) are detected in SN~2017egm with a velocity of $-7000$\,\kms. He~I $\lambda$5876 is the strongest of the four, with a deep absorption feature that could be caused by possible contamination from Na~I $\lambda$5890. The other three lines are weaker, while the He~I $\lambda$7065 absorption is well isolated. Since the four He~I lines are detected in multiple spectra of SN~2017egm between $+95$\,d and $+127$\,d, the existence of helium is firmly established. Further comparisons between SN~2017egm and other known SLSNe-Ib are presented in the next section.

\begin{figure}
\figurenum{9}
\centering
\begin{minipage}{0.5\textwidth}
\centering{\includegraphics[width=1\textwidth]{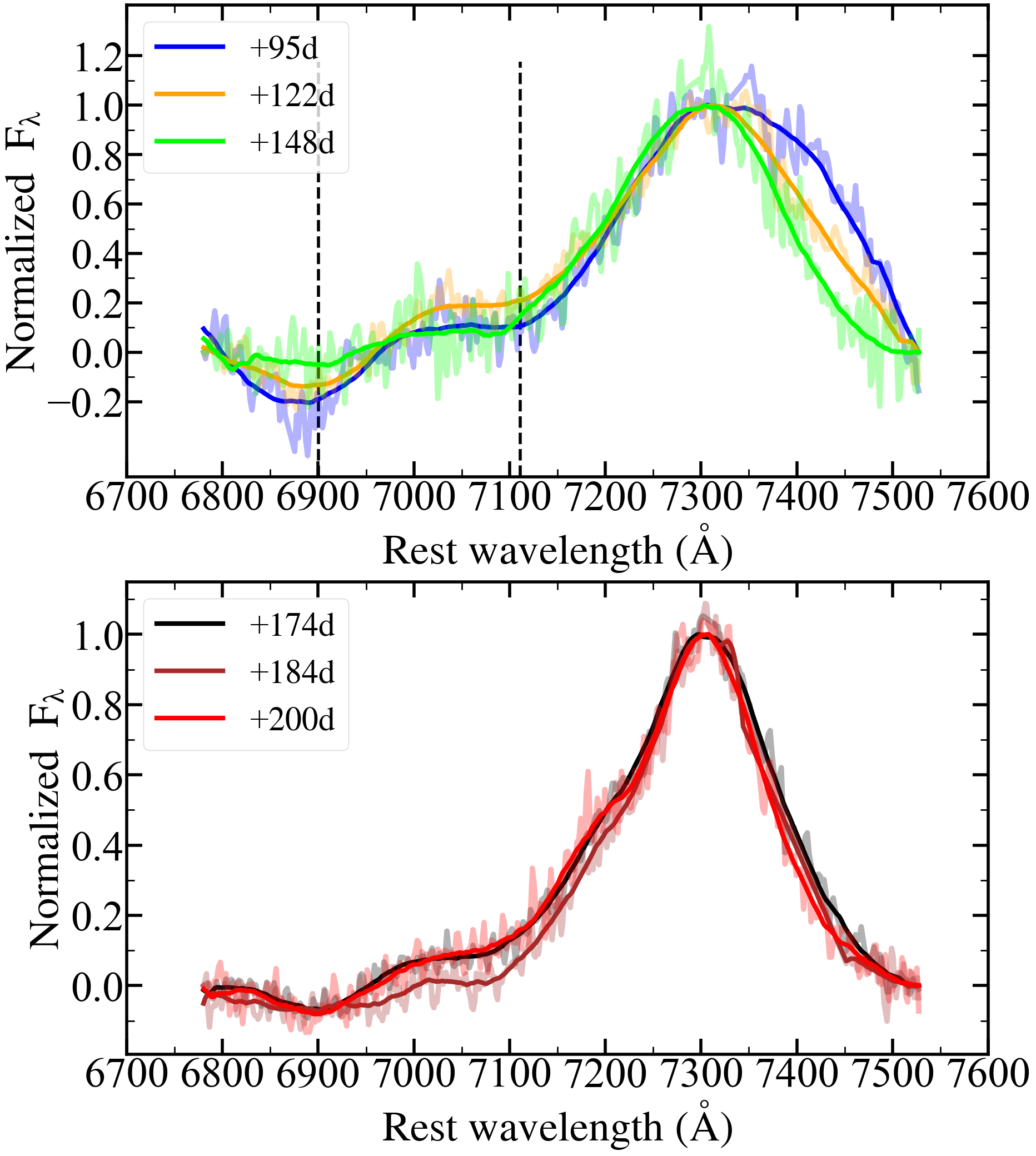}}
\end{minipage}
\caption{The He~I $\lambda\lambda$7065, 7281 evolution of SN~2017egm. All spectra, with continuum subtracted, have been smoothed using the Savitzky-Golay filtering algorithm and normalized to [Ca~II] $\lambda$7300. All spectra are overlaid onto the original spectral data, which are in the same color. The dashed vertical lines mark He~I $\lambda\lambda$7065, 7281, with a velocity of $-7000$\,km\,s$^{-1}$.
 \label{HeI_evolution}}
\end{figure}

\subsection{Spectroscopic Evolution}

Apart from the strong W-shaped O~II absorption features at 3700--4500\,\AA, the early-stage spectra of SN~2017egm are mostly devoid of other prominent features, while heavily blended metallic lines start to appear at $\sim +26$ day (\citealt{Bose2018}; see their Figure 6 for details). Some common line features in SLSNe-Ic are identified, such as a few Fe~II lines and Na~I~D at 4900--5600\,\AA, the Ca~II $\lambda$3750, the Ca~II  $\lambda\lambda$3934, 3969 H\&K doublet, the Ca~II $\lambda\lambda$8498, 8542, 8662 NIR triplet, and O~I $\lambda$7774. At +84\,d, the spectrum is full of strong, blended lines and P~Cygni profiles, with also a strong Mg~I] $\lambda$4571 line, which usually emerges at $\gtrsim 30$ days post-peak (\citealt{Inserra2013, Nicholl2016}). Other prominent features identified are also shown in Figure~\ref{line} and compared with those in SN~2015bn. The Fe~II lines between 4200 and 5000\,\AA, including contributions from Mg~II, become stronger in the high-quality +95\,d spectrum, in which the previously-dominant O~II lines become weaker and finally disappear as the ejecta cool and these species recombine. At the same time, the strengths of [O~I] $\lambda\lambda$6300, 6364, Na~I $\lambda$5890, and the calcium lines increase substantially. The Ca~II $\lambda$3750 is weaker than the prominent Ca~II $\lambda\lambda$3934, 3969 H\&K doublet in both early-time (+26\,d; \citealt{Bose2018}) and late-time (up to +127\,d, see Figure \ref{spec_all}) spectra. There is an unidentified line at $\sim 6100$\,\AA, and a similar line seen for SN~2015bn was interpreted as a possibly detached, high-velocity component of Si~II (\citealt{Nicholl2016}). The asymmetric O~I $\lambda$7774 line profile may be contaminated by Mg~II $\lambda\lambda$7877, 7896. By +148\,d, the blue continuum has faded, and thus the features at the blue end ($< 6000$\,\AA; e.g., Mg~I] $\lambda$4571, Ca~II $\lambda\lambda$3969, 3750, and a few Fe~II lines), which were previously overwhelmed by the continuum, become quite visible. 
It is notable that the inferred velocity of He~I $\lambda$10,830 ($-8000$\,km\,s$^{-1}$) is larger than that of the optical He~I lines ($-7000$\,km\,s$^{-1}$). A similar velocity difference was also found in SN~2019hge (\citealt{Yan2020}; see their Figure 1 for optical He~I lines at $-6000$\,km\,s$^{-1}$ and their Figure 2 for He~I $\lambda$10,830 at $-8000$\,km\,s$^{-1}$), which was attributed to either the difference in excitation levels of the NIR and optical lines or the He~I $\lambda$10,830 being blended with a possible blue shifted C~I $\lambda$10,691 line.
 
The He~I $\lambda$10,830 fades concurrently in the +143\,d NIR spectra (see Figure~\ref{IRspec}) as the He~I $\lambda\lambda$7065, 7281 appears to fade in the  +148\,d optical spectra (see Figure~\ref{HeI_evolution} for the zoomed-in view of the spectra normalized to [Ca~II] $\lambda$7300). Curiously, the apparent fading in He~I happens concurrently when the light curve drops rapidly to a local minimum.
Interestingly, He~I slightly emerges again at +200\,d, coinciding with the second light curve bump. If the He~I features and the light curve indeed co-evolve, it could indicate the source of non-thermal emission needed to produce the He~I features \citep[see, e.g.,][]{Yan2020} may also power the optical emission. We note that the above-mentioned  interpretation is complicated by the possible blending of the [Ca~II] $\lambda$7300 line with oxygen features \citep{Nicholl2019}. We also note that the He~I $\lambda$5876 feature is always present in most optical spectra, though it is heavily blended with Na~I $\lambda$5890.
Most line features can be well fitted by a broad Gaussian and the line velocities decline slowly. For instance, the velocity of [Ca~II] $\lambda$7300 drops from 10,900\,\kms\ (+95\,d) to 8000\,\kms\ (+200\,d).

\begin{figure*}[htb]
\figurenum{10}
\centering
\begin{minipage}{1\textwidth}
\centering{\includegraphics[width=0.95\textwidth]{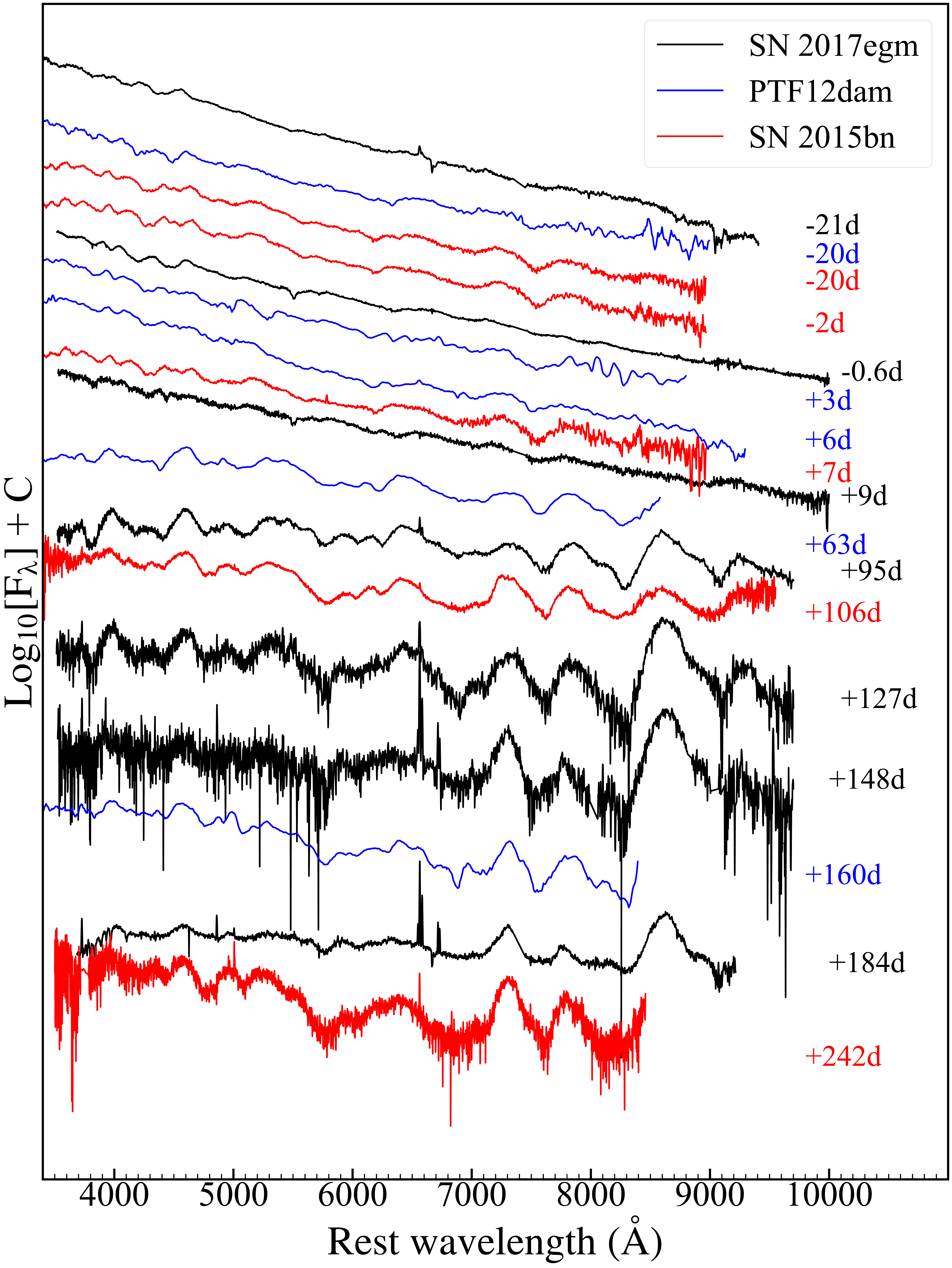}}
\end{minipage}
\caption{The comparison of optical spectra between SN~2017egm and some other slow-declining type Ic SLSNe at similar phases, including PTF12dam (\citealt{Nicholl2013}) and SN 2015bn (\citealt{Nicholl2016}). Their spectroscopic evolution is nearly identical.\label{spec_compare}}
\end{figure*}

The features described above are common for slowly-declining Type Ic SLSNe, so we try to compare the spectral evolution of SN~2017egm with that of PTF12dam and SN~2015bn in Figure~\ref{spec_compare}. However, SN~2017egm differs greatly from the other two objects at early phases; for example, the O~II absorption features disappear faster and the O~I $\lambda$7774 is weaker near the time of peak brightness. Their later evolution is similar except for the presence of strong absorption features at 5876\,\AA\ as well as other weaker He~I lines in SN~2017egm.

\begin{figure*}[htb]
\figurenum{11}
\centering
\begin{minipage}{1\textwidth}
\centering{\includegraphics[width=0.95\textwidth]{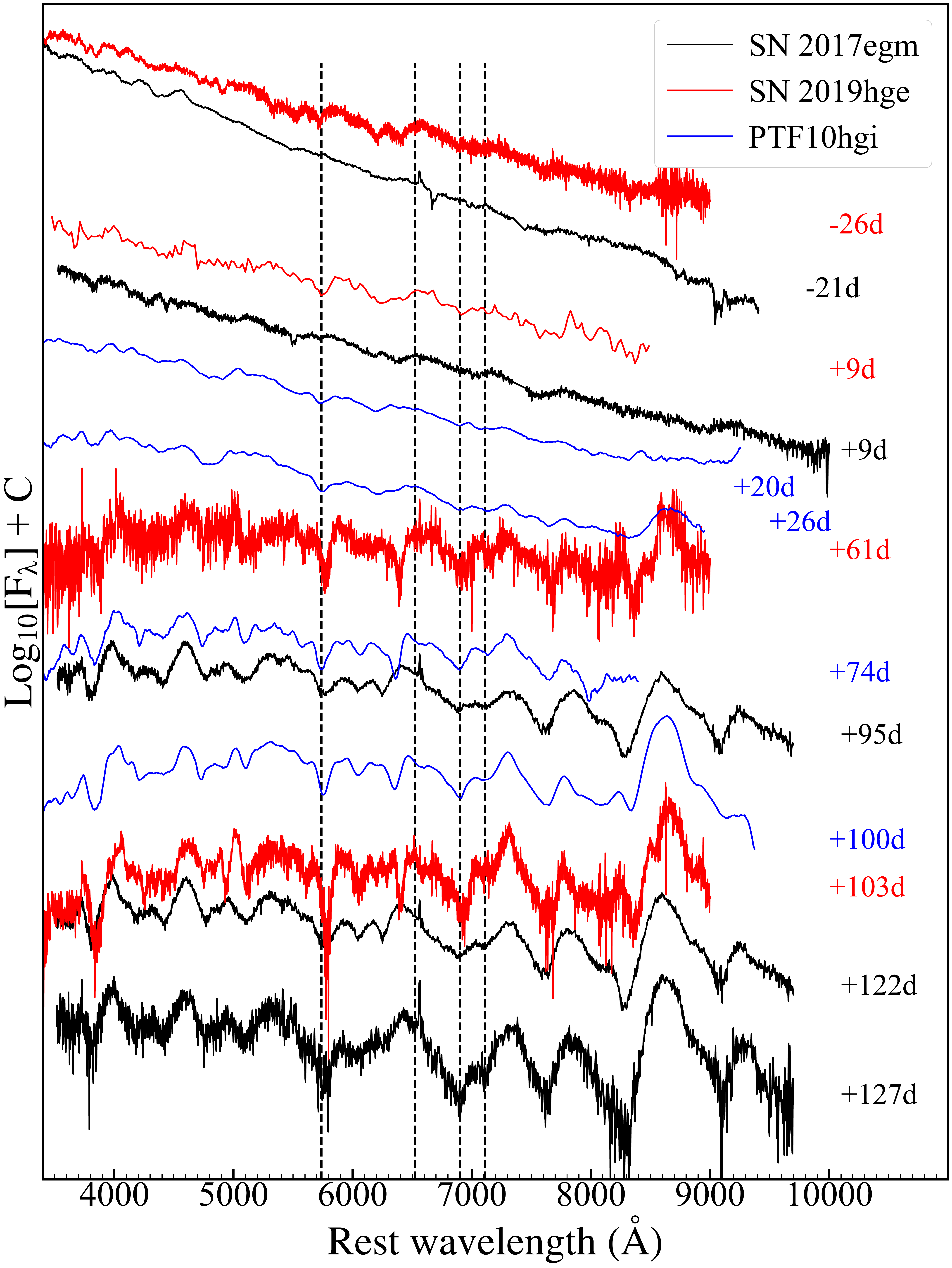}}
\end{minipage}
\caption{Comparison of optical spectra of SN~2017egm and other well-studied SLSNe-Ib/IIb, SN~2019hge~\citep{Yan2020} and PTF10hgi~\citep{Quimby2018}, at similar phases. The dashed lines mark the locations of He~I $\lambda\lambda$5876, 6678, 7065, and 7281, with a velocity of $-7000$\,km\,s$^{-1}$.\label{Ib_compare}}
\end{figure*}

\begin{figure}
\figurenum{12}
\centering
\begin{minipage}{0.5\textwidth}
\centering{\includegraphics[width=1\textwidth]{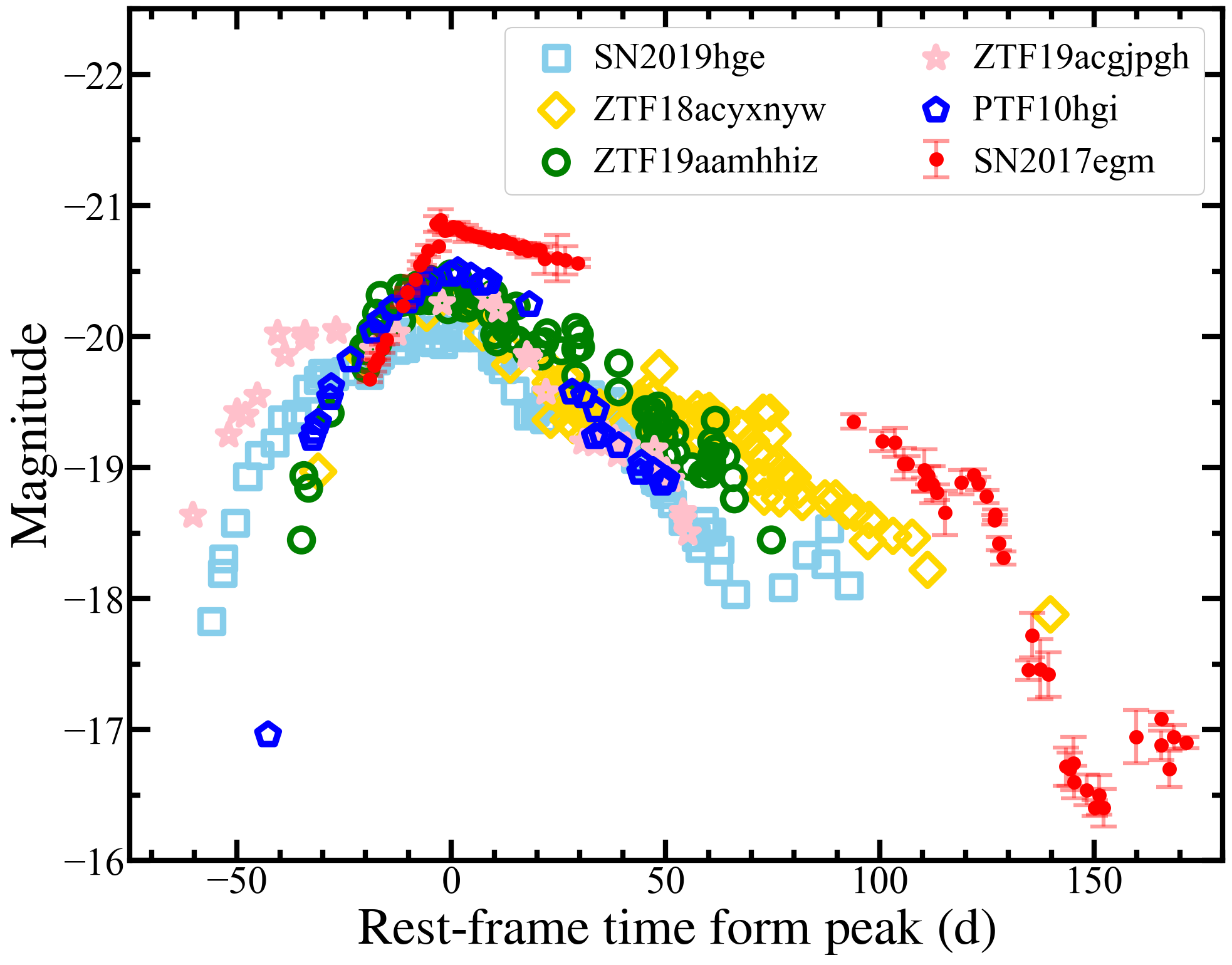}}
\end{minipage}
\caption{The $r$-band light curves of five SLSNe-Ib/IIb with well-sampled observations.
\label{Ib_light curve}}
\end{figure}

\citet{Yan2020} classified SN~2019hge as an SLSN-Ib based on its NIR spectrum and the other five events as good SLSN-Ib/IIb candidates. Here we compare the optical spectra of SN~2017egm with those of SN~2019hge and PTF10hgi (Figure \ref{Ib_compare}). SN~2019hge is redder with strong Si~II $\lambda$6355, C~II $\lambda$6580, and He~I at an early phase, while these features are weaker or absent in SN~2017egm at similar epochs. Similar to SN~2017egm, the He~I in SN~2019hge appeared most prominently when the temperature was relatively low ($T_{\rm bb}\sim 7000-8000$\,K) while suppressed at higher temperature in the earlier phases (\citealt{Yan2020}). 
The He~I absorption features of SN~2017egm, with a velocity of around $-7000$\,km\,s$^{-1}$, are similar with both SN~2019hge and PTF10hgi at late phases. Other nebular-phase features of those objects are similar to those in SLSNe-Ic. Meanwhile, the light curves of the SLSNe-Ib in the ZTF sample (see Figure~\ref{Ib_light curve}) show blue bumps, strong undulations, and mostly relatively low peak luminosities in comparison with SLSNe-Ic~\citep{Chen2022}. Interestingly, with a peak luminosity $\rm M_r \approx -21$\,mag, SN~2017egm is the most luminous among the known SLSNe-Ib, while the peak magnitudes $\rm M_r$ of other SLSN-Ib/IIb are all dimmer than $-20.5$. 
As discussed earlier, the sharp peak of SN~2017egm may suggest a constant-density CSM shell ($s=0$), as previously noted by \citet{Chatzopoulos2013} and \citet{Wheeler2017}.
\citet{Chen2022} also concluded that three of SLSNe-Ib strongly prefer the CSM+Ni model over the magnetar model. However, none of the SLSNe-Ib shows narrow nebular emission line features indicative of ejecta-CSM interaction like in SNe-Ibn or SNe-IIn .

\section{Summary and Discussion}

We have conducted a comprehensive photometric and spectroscopic analysis of SN~2017egm, a hydrogen-poor SLSN that exploded in the rare environment of a massive and metal-rich galaxy, and we identify it as the closest SLSN-Ib found to date. The main results obtained from our analysis are summarized below.

\begin{enumerate}
    \item[$\bullet$] The long-term multiband light curves of SN~2017egm spanning $\sim300$ days have a sharp peak, rapid decline, and multiple late-time bumps, but the object can be broadly categorized into the subclass of slowly-evolving SLSNe-I according to the luminosity and color evolution near peak brightness. 
    
    \item[$\bullet$] The spectra of SN~2017egm resemble those of SLSNe-Ic except for the presence of strong He~I $\lambda$10,830 emission and multiple optical He~I absorption lines. Given the presence of these helium features, we classify SN~2017egm as an SLSN-Ib, which is a rare yet accumulating subpopulation of SLSNe-I \citep{Yan2020}. 
  
    \item[$\bullet$] SN~2017egm is one of a handful of SLSNe-I with a significant IR excess ($L_{\rm IR} \approx 10^8\,\lsun$) indicative of dust echoes from pre-existing or emission from newly-formed dust \citep{Sun2022}.
    
    \item[$\bullet$] SN~2017egm was not detectable in each of the four-epoch {\it Chandra} observations, or even in the stacked images. This yields the tightest constraint on the X-ray emission of an SLSN-I to date, with a $3\sigma$ upper limit of $1.9\times10^{39}$\,\lum\ in the 0.5--10 keV range.

\end{enumerate}

The observational features discussed above serve as the most complete characterization of SN~2017egm to date in terms of the unprecedented temporal and wavelength coverage. Before this work, several models were proposed to account for the observations of SN~2017egm, and its light curve seems more likely to be explained by CSM interaction (\citealt{Wheeler2017, Hosseinzadeh2022}). On the basis of our more complete characterization, we try to further explore this scenario with {\tt{TigerFit}} (\citealt{Chatzopoulos2016, Wheeler2017}) assuming a hybrid model by combining radioactive decay with a constant density (camrads0) or a wind-like density (csmrads2) profile (\citealt{Chatzopoulos2013}). The best-fit  
model has an ejecta mass of 10.7\,\msun\ and a $\rm ^{56}Ni$ mass of 0.15\,\msun, colliding with a CSM with mass of 2.7\,\msun.

\begin{figure}
\figurenum{13}
\centering
\begin{minipage}{0.5\textwidth}
\centering{\includegraphics[width=1\textwidth]{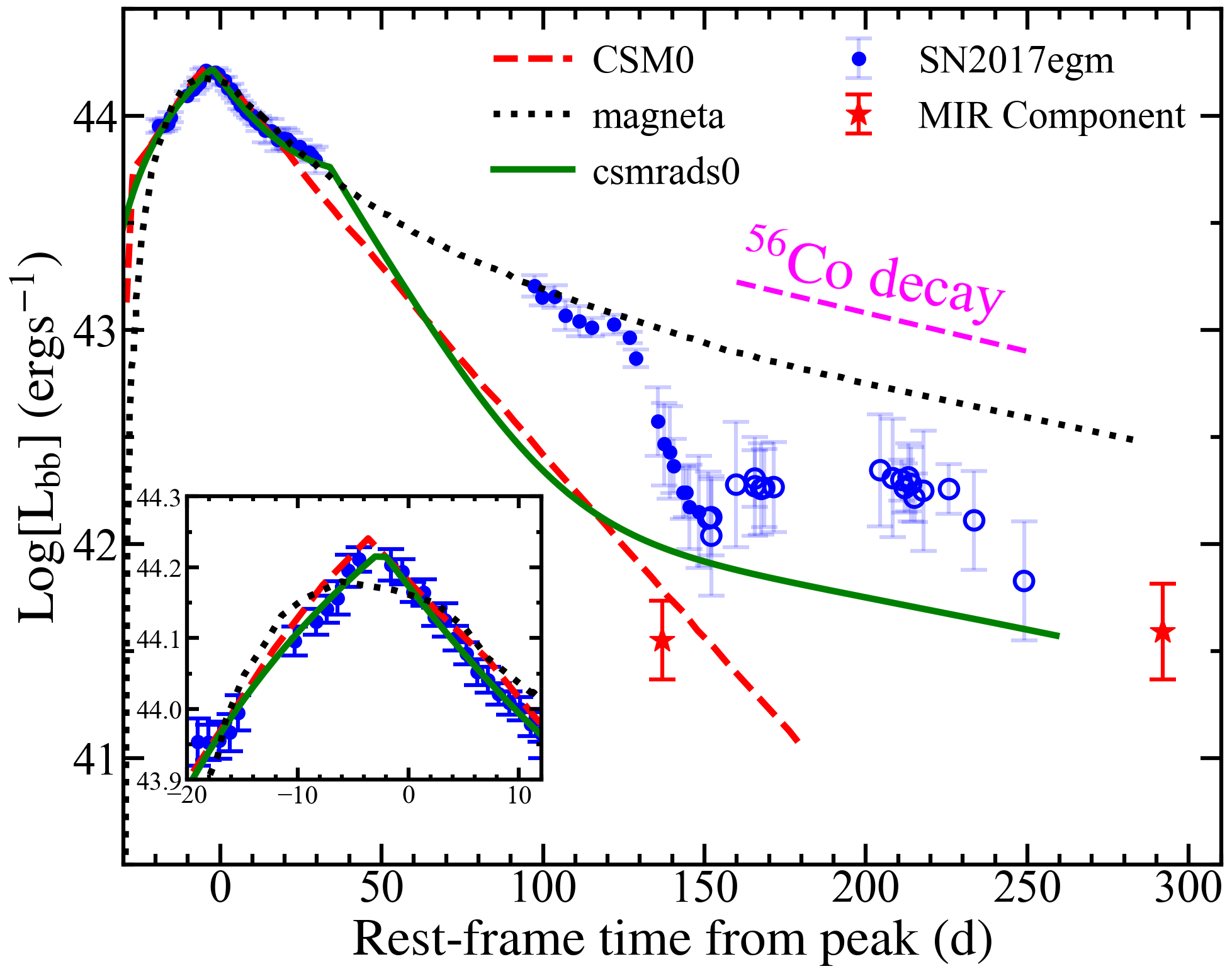}}
\end{minipage}
\caption{The bolometric light curve is plotted together with the best-fit {\tt csmrads0} model and previous CSM0/magnetar model from \citet{Wheeler2017}. The inset shows a zoomed-in view around the peak.\label{model}}
\end{figure}

So far, no simple models, including our best-fit one or others \citep{Wheeler2017, Hosseinzadeh2022}, can reproduce the bolometric light curve of SN~2017egm satisfactorily (see Figure~\ref{model}). The model powered solely by radioactive decay is not viable because it requires substantially more $\rm ^{56}Ni$ than that allowed by the ejecta mass. The magnetar-only model needs a rather smaller ejecta mass, but it fails to reproduce the peak light-curve profile (\citealt{Wheeler2017}; see their Table 1 for the parameters) and late-time fluctuations. The model will release a total energy significantly exceeding the observed value, even counting in the reprocessed energy radiated in MIR which is yet too low to compensate for the sharp optical luminosity drop (see Figure~\ref{model}). The CSM model with a constant density shell could fit well the sharp peak, but a single ejecta-CSM interaction model is inadequate to produce the multiple bumps seen in SN~2017egm.

\citet{Liu2018} has shown that a multiple ejecta-CSM interaction model can fit the undulating bolometric light curve of iPTF15esb, which displays two prominent peaks and a plateau. In their scenario, the appearance of late-time broad \ha\ emission can be naturally interpreted as the ejecta running into a neutral H-shell~\citep{Yan2017b}. Massive stars could experience mass losses in the form of eruptions in the final stage of their lives \citep{Vink2022}, and the progenitors of SNe could sometimes expel two or more shells and/or episodic winds. Furthermore, the steep decline ($\sim 0.1$\,mag\,day$^{-1}$) in SN~2017egm is expected in the CSM-interaction scenario \citep{Sorokina2016}.

Thus, these observations of light curves may indicate the existence of extended and multiple CSM shells or CSM clumps at different radii around SN~2017egm. The ejecta-CSM interactions could be the sources of the non-thermal excitation needed to produce the He~I lines \citep[see, e.g.,][]{Yan2020}, so it might explain why the He~I features appear to co-evolve with the late-time light curve as shown in Figure~\ref{HeI_evolution}. Similar co-evolution of light curve and spectra was seen in the case hydrogen-rich SN~1996al, which was attributed to CSM interation \citep{SN1996al}. 
One major sign from interaction with the CSM is broad emission lines from shocked photoionized ejecta \citep[see, e.g.,][]{handbook}, and the ``hump'' near He~I$\lambda$7065 in +174 -- +200\,d spectra (see the bottom panel of Figure~\ref{HeI_evolution}) may hint the existence of such features blended by the strong Ca~II lines. Detailed modeling will be needed to further model the data and explore the underlying mechanism, but is beyond the scope of this work.

\begin{figure}
\figurenum{14}
\centering
\begin{minipage}{0.5\textwidth}
\centering{\includegraphics[width=1\textwidth]{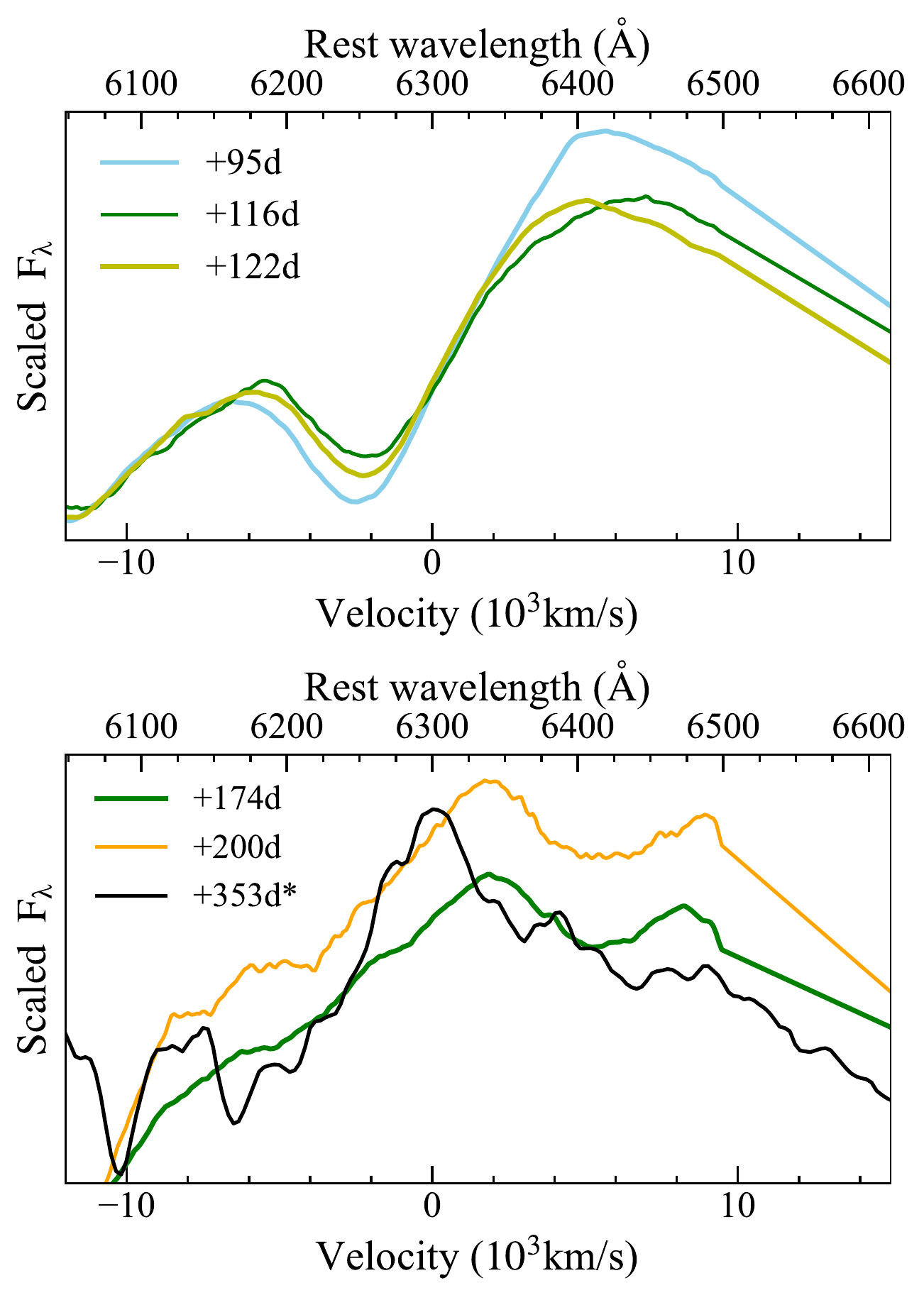}}
\end{minipage}
\caption{Observed [O~I] $\lambda\lambda$6300, 6364 features.
{\it Top panel:} The considerably redshifted features of [O~I] $\lambda\lambda$6300, 6364 before 122\,d.
{\it Bottom panel:} A hinted narrow core near 6300\,\AA\ in late-time spectra was previously reported by \citet{Nicholl2019} (a spectrum at +353\,d from \citealt{Nicholl2019} is marked by $\ast$ in the legend). Our late-time spectra show similar profiles, while the widths of the profiles in our +174\,d and +200\,d spectra are broader than that of the +353\,d spectrum from \citet{Nicholl2019}. The spectra are smoothed using the Savitzky-Golay filtering algorithm and have been continuum-subtracted and converted to velocity space. The strong host-galaxy H$\alpha$ emission was masked.\label{OI}}
\end{figure}

A common issue with the ejecta-CSM interaction interpretation for SLSNe-I is the lack of narrow emission lines expected from the photoionized CSM. A hydrogen-deficient CSM can lead to non-detection of Balmer lines, but the lack of emission from other elements could remains problematic. However, \citet{Chatzopoulos2013} suggested that the absence of narrow emission lines in the optical spectra of SLSNe-I does not necessarily argue against CSM interaction. In that scenario, emission due to CSM interaction may appear as a blue continuum which has been observed in the $\lesssim +200$\,d spectra of SN~2017egm with an H-deficient CSM. They argued that the intermediate-width emission such as [O~I] $\lambda\lambda$6300, 6364 and O~I $\lambda$7774 features found in SN~2007bi could be evidence for CSM interaction \citep{galyam2009}. In fact, \citet{Nicholl2019} noticed a hinted narrow core of [O~I] $\lambda$6300 in the late-time spectra of SN~2017egm, and their latest spectrum at 353\,d had the clearest indication of such a core. Our late-time spectra ($\sim 200$\,d)  show similar profiles (see the bottom panel of Figure~\ref{OI}), albeit larger in width than the profile in their 353\,d spectrum. Interestingly, like the He~I features, the [O~I] feature also changes significantly during the drastic light-curve evolution. Its primary peak was considerably redshifted before +122\,d (see the top panel Figure~\ref{OI}), and the line profiles of [O~I] $\lambda\lambda$6300, 6364 change considerably after the light curve drops rapidly. Since detailed, non-LTE radiation hydrodynamics models of H-poor CSM interaction models are not available, the nature of these lines and their formation sites await further investigation.

We have shown that the color and spectral evolution of SN~2017egm are broadly similar to those of other SLSNe-I such as SN~2015bn and also SLSN-Ib SN~2019hge except for He~I lines, yet they have substantial differences in light-curve evolution. Previous spectroscopic observations of SLSNe-I have not captured the drastic spectral change in the bumpy phase and thus offered us few clues on the light-curve evolution \citep{Hosseinzadeh2022}. However, our high-SNR spectra obtained at different bumpy phases of SN~2017egm suggest clearly that not only the continuum flux variations but also significant changes in spectroscopic profiles, potentially shedding new light on the formation of lines in the ejecta.

In addition, the X-ray and MIR observations may offer useful insights into SN~2017egm. SN~2017egm holds the deepest X-ray upper limit among SLSNe-I, which might provide strict constraints on the subparsec environment and properties of central engines in SLSNe-I \citep[see, e.g.][]{Chevalier2003, Levan2013, Margutti2018, Huang2018, Moriya2018, Vurm2021}. The tightest X-ray observations in SLSNe-I rule out the densest environments typical of luminous blue variable eruptions and Type IIn SNe (\citealt{Margutti2018}), but the parameter space of SN~2017egm ($L_{\rm X}<10^{40}$\,erg\,s$^{-1}$ and X-ray-to-optical luminosity ratio $L_{\rm X}/L_{\rm Opt}\lesssim 10^{-3}$) is almost an entirely uncharted territory to explore. \citet{AndrewsSmith2018} invoked a possibility that strong CSM interaction is hidden behind the ejecta over a wide range of viewing angles to explain the absence of narrow emission lines and X-ray emissions in the case of the peculiar iPTF14hls, and similar scenarios may be of interest to consider for SLSNe-Ib in the context of CSM interactions. On the other hand, the detection of an unambiguous dust emission has not been reported in SLSNe until very recently \citep{Chen2021, Sun2022}, and SN~2017egm belongs to the population with clear evidence of pre-existing dust or dust formation. The MIR echo agrees nicely with the ejecta-CSM interaction model, and reveals a new promising method to probe the subparsec environment of SLSNe. 

In the upcoming golden era of wide-field time-domain surveys, such as the Legacy Survey of Space and Time (LSST; \citealt{Ivezic2019}) with the Rubin Observatory and the Wide-Field Survey Telescope (WFST; \citealt{Lin2022}), SLSNe are expected to soon be discovered at a high rate. SN~2017egm, as a rare SLSN studied in great detail, will serve as a valuable case for comparison and be re-evaluated statistically in the context of large samples.

\acknowledgements

We thank the anonymous referee for carefully reading our manuscript and providing valuable comments. This work is supported by the National Natural Science Foundation of China (grants 11833007, 12133005, 12073025, 12192221) and the Fundamental Research Funds for the Central Universities (WK3440000006). S.D. acknowledges support from the Xplorer Prize. J.Z and N.J. gratefully acknowledge the support of Cyrus Chun Ying Tang Foundations. S.M. acknowledges support from the Magnus Ehrnrooth Foundation and the Vilho, Yrj\"{o}, and Kalle V\"{a}is\"{a}l\"{a} Foundation. A.P., N.E.R., S.B., and E.C. are partially supported by the PRIN-INAF 2022 project ''Shedding light on the nature of gap transients: from the observations to the models''. A.V.F.'s supernova group at UC Berkeley is grateful for financial support from the Christopher R. Redlich Fund and numerous individual donors. S.M. acknowledges support from the Academy of Finland project 350458. 

Based in part on observations made with the Nordic Optical Telescope, owned in collaboration by the University of Turku and Aarhus University, and operated jointly by Aarhus University, the University of Turku, and the University of Oslo (respectively representing Denmark, Finland, and Norway), the University of Iceland, and Stockholm University, at the Observatorio del Roque de los Muchachos, La Palma, Spain, of the Instituto de Astrofisica de Canarias.

This research uses data obtained through the Telescope Access Program (TAP). Observations with the Hale telescope at Palomar Observatory were obtained as part of an agreement between the National Astronomical Observatories, Chinese Academy of Sciences, and the California Institute of Technology. We thank the {\it Swift} PI Brad Cenko, the Observation Duty Scientists, and the science planners for approving and executing our {\it Swift}/UVOT observations. 

A major upgrade of the Kast spectrograph on the Shane 3\,m telescope at Lick Observatory, led by Brad Holden, was made possible through generous gifts from the Heising-Simons Foundation, William and Marina Kast, and the University of California Observatories. Research at Lick Observatory is partially supported by a generous gift from Google.
This work is partly based on observations made with the Gran Telescopio Canarias (GTC), installed at the Spanish Observatorio del Roque de los Muchachos of the Instituto de Astrofísica de Canarias, on the island of La Palma.

The LBT is an international collaboration among institutions in the United States, Italy, and Germany. LBT Corporation partners are The University of Arizona on behalf of the Arizona Board of Regents; Istituto Nazionale di Astrofisica, Italy; LBT Beteiligungsgesellschaft, Germany, representing the Max-Planck Society, The Leibniz Institute for Astrophysics Potsdam, and Heidelberg University; The Ohio State University, representing OSU, University of Notre Dame, University of Minnesota, and University of Virginia. This paper made use of the modsCCDRed data reduction code developed in part with funds provided by NSF grants AST-9987045 and AST-1108693. This paper made use of the modsIDL spectral data reduction pipeline developed in part with 
funds provided by NSF grant AST-1108693 and a generous gift from OSU Astronomy alumnus David G. Price through the Price Fellowship in Astronomical Instrumentation.


\begin{deluxetable*}{c c c c c c c c c c}
\tablenum{1}

\tablecaption{Photometry of SN~2017egm.\label{photometry_table}}
\tablehead{ MJD & Phase$^{a}$ & B & g & V & r & i & z & Telescope$^{b}$ & Source$^{d}$ \\
       & (days) & (mag) & (mag) & (mag) & (mag) & (mag) & (mag) & /Inst. &}
\startdata
57904.90 & -20.28 & ... & ... & ... & ... & ... & 16.74 $\pm$ 0.04 & LT & Bose et al. 2018\\
57905.18 & -20.01 & ... & ... & 15.89 $\pm$ 0.02 & ... & ... & ... & PO & Bose et al. 2018 \\
57906.20 & -19.02 & 15.82 $\pm$ 0.04 & 15.76 $\pm$ 0.03 & 15.81 $\pm$ 0.02 & 16.03 $\pm$ 0.02 & 16.21 $\pm$ 0.03 & ... & PO & Bose et al. 2018\\
57907.20 & -18.05 & 15.81 $\pm$ 0.02 & 15.76 $\pm$ 0.03 & 15.77 $\pm$ 0.03 & 15.93 $\pm$ 0.05 & 16.12 $\pm$ 0.03 &...& PO & Bose et al. 2018\\
57908.19 & -17.09 &...& 15.55 $\pm$ 0.01 &...&15.88 $\pm$ 0.01&16.06 $\pm$ 0.02&...& 1.0m LCOGT&Hosseinzadeh et al. 2022\\
57908.20 & -17.08 & 15.76 $\pm$ 0.04&15.67 $\pm$ 0.04 & 15.71 $\pm$ 0.03 & 15.88 $\pm$ 0.05 & 16.08 $\pm$ 0.03 & ...& PO &Bose et al. 2018\\
57909.20 & -16.11 & ... & 15.58 $\pm$ 0.02 & 15.64 $\pm$ 0.02 & 15.80 $\pm$ 0.03 & 16.00 $\pm$ 0.03 & ...& PO &Bose et al. 2018\\
57909.94 & -15.39 &...&...&...&...&...&16.40 $\pm$ 0.05& LT & Bose et al. 2018\\
57910.20 & -15.14 &15.67 $\pm$ 0.04&15.53 $\pm$ 0.05&15.58 $\pm$ 0.02&15.73 $\pm$ 0.03& 15.94 $\pm$ 0.03 &...& PO & Bose et al. 2018\\
57911.22 & -14.15 & 15.60 $\pm$ 0.03&...&15.51 $\pm$ 0.02&...&15.86 $\pm$ 0.03&...&PO&Bose et al. 2018\\
57914.18 & -11.27 &...& 15.21 $\pm$ 0.01&...&15.48 $\pm$ 0.01 & 15.60 $\pm$ 0.01& ...& 1.0m LCOGT &Hosseinzadeh et al. 2022\\
57914.20 & -11.26 & 15.32 $\pm$ 0.03&...&15.29 $\pm$ 0.02&...&15.64 $\pm$ 0.03 &...& PO& Bose et al. 2018\\
57915.15 & -10.33 &...& 15.13 $\pm$ 0.02&...&15.38 $\pm$ 0.02 & 15.59 $\pm$ 0.02& ...& 1.0m LCOGT &Hosseinzadeh et al. 2022\\
57915.20 & -10.29 & 15.26 $\pm$ 0.03&15.21 $\pm$ 0.03& 15.20 $\pm$ 0.03 & 15.37 $\pm$ 0.03& 15.57 $\pm$ 0.03 &...&PO&Bose et al. 2018\\
57915.90 & -9.61 &...&...&...&...&...&15.95 $\pm$ 0.03& LT &Bose et al. 2018\\
57917.20 & -8.35 & ...& 15.02 $\pm$ 0.02& 15.08 $\pm$ 0.04& 15.28 $\pm$ 0.03& 15.43 $\pm$ 0.04&...&PO&Bose et al. 2018\\
57918.20 & -7.38 & 15.06 $\pm$ 0.02 & 14.95 $\pm$ 0.04&14.97 $\pm$ 0.02& 15.16 $\pm$ 0.03& 15.37 $\pm$ 0.03&...&PO&Bose et al. 2018\\ 
57919.21 & -6.40 & 14.96 $\pm$ 0.03& 14.90 $\pm$ 0.03& 14.93 $\pm$ 0.03& 15.12 $\pm$ 0.04 & 15.32 $\pm$ 0.03&...& PO& Bose et al. 2018\\ 
57920.22 & -5.42 & 14.94 $\pm$ 0.03&14.88 $\pm$ 0.03& 14.88 $\pm$ 0.03&15.06 $\pm$ 0.05& 15.25 $\pm$ 0.03&...& PO & Bose et al.2018\\
57921.22 & -4.45 & 14.85 $\pm$ 0.04& 14.81 $\pm$ 0.04& 14.88 $\pm$ 0.03& ...&...&...&PO&Bose et al. 2018\\
57922.15 & -3.54 &...&14.57 $\pm$ 0.07&...&14.85 $\pm$ 0.06&15.06 $\pm$ 0.08&...& KeplerCam& Hosseinzadeh et al. 2022\\
57922.71 & -3.00 &...&14.65 $\pm$ 0.04&...&15.02 $\pm$ 0.04&15.19 $\pm$ 0.04&...& Templeton& Hosseinzadeh et al. 2022\\
57923.15 & -2.57 &...&14.57 $\pm$ 0.14&...&14.82 $\pm$ 0.08&15.02 $\pm$ 0.08&...& KeplerCam& Hosseinzadeh et al. 2022\\
57923.91 & -1.83 &...&...&...&...&...&15.61 $\pm$ 0.05&AF&Bose et al. 2018\\
57924.20 & -1.56 & 14.80 $\pm$ 0.03& 14.71 $\pm$ 0.03& 14.71 $\pm$ 0.02& 14.90 $\pm$ 0.04& 15.09 $\pm$ 0.03&...&PO&Bose et al. 2018\\
57925.22 & -0.57 & 14.78 $\pm$ 0.04 & 14.71 $\pm$ 0.04& 14.73 $\pm$ 0.03& 14.89 $\pm$ 0.02 &15.08 $\pm$ 0.03&...&PO&Bose et al.2018\\
57926.22 & 0.40 & 14.80 $\pm$ 0.01 & 14.70 $\pm$ 0.02&14.73 $\pm$ 0.03&14.87 $\pm$ 0.02& 15.04 $\pm$ 0.05&...&PO&Bose et al.2018\\
57927.22 & 1.37 &14.83 $\pm$ 0.04& 14.73 $\pm$ 0.05& 14.71 $\pm$ 0.02&14.88 $\pm$ 0.04& 15.07 $\pm$ 0.03&...&PO&Bose et al. 2018\\
57928.21 & 2.34 & 14.86 $\pm$ 0.01&14.77 $\pm$ 0.03&14.69 $\pm$ 0.06&14.91 $\pm$ 0.03&15.08 $\pm$ 0.04&...&PO&Bose et al. 2018\\
57928.92 & 3.03 &...&...&...&...&...&15.49 $\pm$ 0.05&AF&Bose et al. 2018\\
57929.22 & 3.31 & 14.93 $\pm$ 0.09&14.80 $\pm$ 0.04&14.77 $\pm$ 0.05&14.92 $\pm$ 0.06&15.09 $\pm$ 0.09&...&PO&Bose et al. 2018\\
57930.21 & 4.28 &14.94 $\pm$ 0.02&14.80 $\pm$ 0.05&14.78 $\pm$ 0.03&14.93 $\pm$ 0.04&15.08 $\pm$ 0.03 &...&PO&Bose et al. 2018\\
57931.22 & 5.26 &14.94 $\pm$ 0.01&14.82 $\pm$ 0.03&14.79 $\pm$ 0.02&14.94 $\pm$ 0.03&15.09 $\pm$ 0.03&...&PO&Bose et al. 2018\\
57932.21 & 6.22 &14.99 $\pm$ 0.02&14.84 $\pm$ 0.02&14.83 $\pm$ 0.04&14.95 $\pm$ 0.03& 15.09 $\pm$ 0.03&...&PO&Bose et al. 2018\\
57933.21 & 7.19 &15.01 $\pm$ 0.03&14.84 $\pm$ 0.03&14.83 $\pm$ 0.03&14.95 $\pm$ 0.02&15.10 $\pm$ 0.04&...&PO&Bose et al. 2018\\
57934.21 & 8.16 & 15.03 $\pm$ 0.01&14.85 $\pm$ 0.02&14.86 $\pm$ 0.02&14.96 $\pm$ 0.02&15.11 $\pm$ 0.03&...&PO&Bose et al. 2018\\
57935.21 & 9.13 & 15.03 $\pm$ 0.01&14.88 $\pm$ 0.02&14.88 $\pm$ 0.03&14.98 $\pm$ 0.03&15.11 $\pm$ 0.03&...&PO&Bose et al. 2018\\
57935.89 & 9.79 &...&...&...&...&...&15.42 $\pm$ 0.05&AF&Bose et al. 2018\\
57936.21 & 10.10 & 15.06 $\pm$ 0.02&14.90 $\pm$ 0.03&14.88 $\pm$ 0.02&14.97 $\pm$ 0.03&15.11 $\pm$ 0.04&...&PO&Bose et al. 2018\\
57937.21 & 11.07 & 15.09 $\pm$ 0.01&14.92 $\pm$ 0.04&14.90 $\pm$ 0.04&14.99 $\pm$ 0.03&15.12 $\pm$ 0.05&...&PO&Bose et al. 2018\\
57938.21 & 12.04 & 15.09 $\pm$ 0.01&14.94 $\pm$ 0.03&14.92 $\pm$ 0.02&14.97 $\pm$ 0.02&15.12 $\pm$ 0.04&...&PO&Bose et al. 2018\\
57939.21 & 13.01 & 15.05 $\pm$ 0.01&14.95 $\pm$ 0.04&14.93 $\pm$ 0.04&14.99 $\pm$ 0.03&15.11 $\pm$ 0.06&...&PO&Bose et al. 2018\\
57940.21 & 13.98 & 15.09 $\pm$ 0.03&14.97 $\pm$ 0.02&14.98 $\pm$ 0.04&15.00 $\pm$ 0.03&15.14 $\pm$ 0.05&...&PO&Bose et al. 2018\\
57941.21 & 14.95 & 15.15 $\pm$ 0.03&...&15.05 $\pm$ 0.03&...&...&...&PO&Bose et al. 2018\\
57942.20 & 15.91 & 15.13 $\pm$ 0.03&14.99 $\pm$ 0.05&15.01 $\pm$ 0.03&15.04 $\pm$ 0.04&15.16 $\pm$ 0.04&...&PO&Bose et al. 2018\\
57943.20 & 16.88 & 15.17 $\pm$ 0.02&...&...&15.02 $\pm$ 0.03&15.18 $\pm$ 0.05&...&PO&Bose et al. 2018\\
57944.20 & 17.85 & 15.19 $\pm$ 0.06&15.05 $\pm$ 0.03&15.01 $\pm$ 0.05&15.05 $\pm$ 0.05&15.16 $\pm$ 0.04&...&PO&Bose et al. 2018\\
57946.20 & 19.79 & 15.21 $\pm$ 0.01 &15.08 $\pm$ 0.03&15.01 $\pm$ 0.05&15.05 $\pm$ 0.03&15.21 $\pm$ 0.11&...&PO&This work\\
57947.20 & 20.76 & ...&15.05 $\pm$ 0.02&15.06 $\pm$ 0.04&15.05 $\pm$ 0.02&15.21 $\pm$ 0.07&...&PO&This work\\
57948.20 & 21.73 & ...&15.09 $\pm$ 0.02&15.05 $\pm$ 0.03&15.12 $\pm$ 0.11&15.25 $\pm$ 0.11&...&PO&This work\\
57951.21 & 24.65 & 15.26 $\pm$ 0.01&15.14 $\pm$ 0.02&15.13 $\pm$ 0.03&15.11 $\pm$ 0.18&15.18 $\pm$ 0.08&...&PO&This work\\
\enddata      
\end{deluxetable*}

\begin{deluxetable*}{c c c c c c c c c c}
\tablenum{1}

\tablecaption{\it Continued}
\tablehead{ MJD & Phase$^{a}$ & B & g & V & r & i & z & Telescope$^{b}$ & Source \\
       & (days) & (mag) & (mag) & (mag) & (mag) & (mag) & (mag) & /Inst. &}
\startdata
57952.90 & 26.29 &...&...&...&...&...&15.44 $\pm$ 0.05&AF&This work\\
57953.20 & 26.59 & ...&...&15.04 $\pm$ 0.05&15.13 $\pm$ 0.11&...&...&PO&This work\\
57954.20 & 27.56 & 15.30 $\pm$ 0.04&15.16 $\pm$ 0.03&...&...&...&...&PO&This work\\
57955.20 & 28.53 & ...&15.22 $\pm$ 0.03&15.04 $\pm$ 0.05&...&...&...&PO&This work\\
57956.20 & 29.50 & 15.37 $\pm$ 0.03&...&...&15.15 $\pm$ 0.03&...&...&PO&This work\\
57957.20 & 30.47 & ...& 15.30 $\pm$ 0.07 &...&...&...&...&PO&This work\\
57958.20 & 31.44 & ...&...& 15.09 $\pm$ 0.04 &...&...&...&PO&This work\\
58012.50 & 84.12 & ...&16.25 $\pm$ 0.09&...&...&...&...&PO&This work\\
58018.50 & 89.94 & ...&...&16.30 $\pm$ 0.10&...&...&...&PO&This work\\
58019.50 & 90.92 & ...&...&16.16 $\pm$ 0.17&...&...&...&PO&This work\\
58023.50 & 94.80 & ...&...&...&...&16.23 $\pm$ 0.17&...&PO&This work\\
58026.50 & 97.71 & ...&...&16.54 $\pm$ 0.05&...&...&...&PO&This work\\
58028.50 & 99.65 & 17.22 $\pm$ 0.03&16.72 $\pm$ 0.21&...&...&16.48 $\pm$ 0.07&16.30 $\pm$ 0.07&PO, LT&This work\\
58028.90 & 100.04 & ...&...&...&...&...&16.34 $\pm$ 0.04&LT&This work\\
58029.25 & 100.38& ...&...&16.52 $\pm$ 0.07&16.51 $\pm$ 0.08&16.65 $\pm$ 0.10&16.34 $\pm$ 0.10&PO, LT&This work\\
58029.90 & 101.01 &...&...&...&...&...&16.28 $\pm$ 0.51&LT&This work\\
58030.50 & 101.59 & ...&...&...&...&16.33 $\pm$ 0.13 &...&LT&This work\\
58032.52 & 103.55 & ...&16.86 $\pm$ 0.15&...&16.52 $\pm$ 0.11&16.54 $\pm$ 0.17&...&KeplerCam&Hosseinzadeh et al. 2022\\
58033.50 & 104.50 & ...&...&...&...&16.51 $\pm$ 0.04 &...&LT&This work\\
58034.50 & 105.47 & ...&...&16.45 $\pm$ 0.14&...&...&...&PO&This work\\
58034.52 & 105.49 & ...&16.94 $\pm$ 0.11&...&16.68 $\pm$ 0.12&...&...&KeplerCam&Hosseinzadeh et al. 2022\\
58035.20 & 106.15 & 17.51 $\pm$ 0.06&...&16.66 $\pm$ 0.10&...&16.66 $\pm$ 0.07&16.47 $\pm$ 0.09&LT&This work\\
58035.52 & 106.46 & ...&16.99 $\pm$ 0.15&...&...&...&...&KeplerCam&Hosseinzadeh et al. 2022\\
58035.90 & 106.83 & 17.43 $\pm$ 0.10&16.81 $\pm$ 0.19&16.64 $\pm$ 0.10&16.68 $\pm$ 0.06&...&16.46 $\pm$ 0.21&LT&This work\\
58039.50 & 110.32 & 17.56 $\pm$ 0.05&...&16.82 $\pm$ 0.07&16.84 $\pm$ 0.05&16.78 $\pm$ 0.07&...&PO&This work\\
58039.53 & 110.35 & ...&...&...&16.73 $\pm$ 0.16&...&...&KeplerCam&Hosseinzadeh et al. 2022\\
58040.51 & 111.30 & ...&17.05 $\pm$ 0.08&...&16.77 $\pm$ 0.07&16.72 $\pm$ 0.15&...&KeplerCam&Hosseinzadeh et al. 2022\\
58041.51 & 112.27 & ...&17.18 $\pm$ 0.10&...&16.84 $\pm$ 0.14&16.78 $\pm$ 0.08&...&KeplerCam&Hosseinzadeh et al. 2022\\
58042.51 & 113.24 & ...&17.11 $\pm$ 0.08&...&16.90 $\pm$ 0.06&16.85 $\pm$ 0.06&...&KeplerCam&Hosseinzadeh et al. 2022\\
58044.50 & 115.17 & 17.68 $\pm$ 0.10&17.04 $\pm$ 0.22&16.98 $\pm$ 0.03&17.05 $\pm$ 0.17&16.88 $\pm$ 0.04&16.61 $\pm$ 0.14&PO, LT&This work\\
58048.50 & 119.05 & 17.63 $\pm$ 0.07&...&16.82 $\pm$ 0.05&16.82 $\pm$ 0.09&...&...&PO&This work\\
58050.16 & 120.66 & 17.56 $\pm$ 0.12&16.94 $\pm$ 0.26&...&...&16.79 $\pm$ 0.08&16.58 $\pm$ 0.11&LT&This work\\
58051.50 & 121.96 & 17.62 $\pm$ 0.02&...&16.79 $\pm$ 0.06&16.77 $\pm$ 0.04&16.76 $\pm$ 0.04&...&PO&This work\\
58052.52 & 122.95 & ...&16.93 $\pm$ 0.06&...&16.83 $\pm$ 0.05&...&...&KeplerCam&Hosseinzadeh et al. 2022\\
58054.52 & 124.88 & ...&16.97 $\pm$ 0.04&...&16.93 $\pm$ 0.05&...&...&KeplerCam&Hosseinzadeh et al. 2022\\
58056.50 & 126.82 & 17.82 $\pm$ 0.04&17.17 $\pm$ 0.19&17.04 $\pm$ 0.02&17.11 $\pm$ 0.04& 17.09 $\pm$ 0.03&16.71 $\pm$ 0.14&PO, LT&This work\\
58056.51 & 126.83 & ...&17.25 $\pm$ 0.04&...&17.07 $\pm$ 0.04&17.04 $\pm$ 0.06&...&KeplerCam&Hosseinzadeh et al. 2022\\
58057.46 & 127.75 & ...&17.30 $\pm$ 0.05&...&17.29 $\pm$ 0.05&...&...&KeplerCam&Hosseinzadeh et al. 2022\\
58058.49 & 128.75 & ...&17.42 $\pm$ 0.04&...&17.40 $\pm$ 0.05&17.27 $\pm$ 0.06&...&KeplerCam&Hosseinzadeh et al. 2022\\
58064.50 & 134.58 & 18.76 $\pm$ 0.08&...&17.81 $\pm$ 0.12&18.26 $\pm$ 0.07&...&...&PO&This work\\
58065.45 & 135.50 & ...&18.13 $\pm$ 0.15&...&17.99 $\pm$ 0.17&...&...&KeplerCam&Hosseinzadeh et al. 2022\\
58067.46 & 137.45 & ...&18.53 $\pm$ 0.09&...&18.25 $\pm$ 0.23&18.18 $\pm$ 0.22&...&KeplerCam&Hosseinzadeh et al. 2022\\
58068.26 & 138.23 & 19.16 $\pm$ 0.25&18.43 $\pm$ 0.07&18.86 $\pm$ 0.49&...&...&17.98 $\pm$ 0.47&LT&This work\\
58069.42 & 139.35 & ...&18.63 $\pm$ 0.14&...&18.29 $\pm$ 0.17&18.12 $\pm$ 0.16&...&KeplerCam&Hosseinzadeh et al. 2022\\
58070.51 & 140.41 & ...&18.76 $\pm$ 0.11&...&...&18.32 $\pm$ 0.08&...&KeplerCam&Hosseinzadeh et al. 2022\\
58072.70 & 142.53 & ...&18.56 $\pm$ 0.44&...&...&...&18.36 $\pm$ 0.13&LT&This work\\
58073.15 & 142.98 & 20.13 $\pm$ 0.14&18.72 $\pm$ 0.16&...&...&19.02 $\pm$ 0.14&18.25 $\pm$ 0.13&LT&This work\\
58073.60 & 143.41 & 19.85 $\pm$ 0.26&19.23 $\pm$ 0.27&19.08 $\pm$ 0.20&18.99 $\pm$ 0.14&18.78 $\pm$ 0.15&18.43 $\pm$ 0.09&2.0m LCOGT&This work\\
58074.60 & 144.38 & 19.87 $\pm$ 0.36&19.11 $\pm$ 0.18&19.20 $\pm$ 0.18&19.01 $\pm$ 0.13&18.95 $\pm$ 0.17&18.23 $\pm$ 0.17&2.0m LCOGT&This work\\
58075.50 & 145.25 & 19.83 $\pm$ 0.31&...&19.01 $\pm$ 0.33&18.97 $\pm$ 0.20&18.68 $\pm$ 0.22&...&PO&This work\\
58075.60 & 145.35 & 20.19 $\pm$ 0.24&19.30 $\pm$ 0.18&19.27 $\pm$ 0.14&19.11 $\pm$ 0.12&18.88 $\pm$ 0.17&18.32 $\pm$ 0.17&2.0m LCOGT&This work\\
\enddata      
\end{deluxetable*}

\begin{deluxetable*}{c c c c c c c c c c}
\tablenum{1}

\tablecaption{\it Continued}
\tablehead{ MJD & Phase$^{a}$ & B & g & V & r & i & z & Telescope$^{b}$ & Source \\
       & (days) & (mag) & (mag) & (mag) & (mag) & (mag) & (mag) & /Inst. &}
\startdata
58078.60 & 148.26 & 20.45 $\pm$ 0.25&19.38 $\pm$ 0.18&19.59 $\pm$ 0.14&19.17 $\pm$ 0.12& 19.03 $\pm$ 0.29&18.45 $\pm$ 0.16&2.0m LCOGT&This work\\
58080.50 & 150.10 & ...&...&19.35 $\pm$ 0.08&19.31 $\pm$ 0.06&...&...&PO&This work\\
58081.20 & 150.78 & 20.43 $\pm$ 0.32&...&19.43 $\pm$ 0.18&...&...&...&LT&This work\\
58081.60 & 151.17 & 20.20 $\pm$ 0.38&19.48 $\pm$ 0.23&19.55 $\pm$ 0.24&19.21 $\pm$ 0.15&19.15 $\pm$ 0.23&18.53 $\pm$ 0.21&2.0m LCOGT&This work\\
58082.59 & 152.13 & 20.38 $\pm$ 0.62&19.45 $\pm$ 0.30&19.59 $\pm$ 0.21&19.31 $\pm$ 0.14&19.13 $\pm$ 0.18&18.38 $\pm$ 0.14&2.0m LCOGT&This work\\
58090.55 & 159.85 &...&19.08 $\pm$ 0.32&18.97 $\pm$ 0.36&18.77 $\pm$ 0.20&18.52 $\pm$ 0.16&18.00 $\pm$ 0.13&2.0m LCOGT&This work\\
58094.22 & 163.41 &...&...&...&...&...&18.20 $\pm$ 0.07&LT&This work\\
58096.50 & 165.63 &19.50 $\pm$ 0.10&...&...&18.63 $\pm$ 0.05&18.70 $\pm$ 0.09&...&PO&This work\\
58096.60 & 165.72 &...&...&...&18.83 $\pm$ 0.11& 18.83 $\pm$ 0.11&...&2.0m LCOGT&This work\\
58098.50 & 167.57 &19.96 $\pm$ 0.25&...&19.01 $\pm$ 0.14&18.67 $\pm$ 0.13&...&...&PO, 2.0m LCOGT&This work\\
58099.50 & 168.54 &...&19.00 $\pm$ 0.11&...&18.77 $\pm$ 0.09&18.75 $\pm$ 0.18&17.90 $\pm$ 0.14&2.0m LCOGT&This work\\
58102.50 & 171.45 &20.09 $\pm$ 0.10&...&19.06 $\pm$ 0.08&18.81 $\pm$ 0.04&18.88 $\pm$ 0.11&...&PO&This work\\
58102.55 & 171.50 &...&18.88 $\pm$ 0.12&...&...&18.64 $\pm$ 0.20&...&KeplerCam&Hosseinzadeh et al. 2022\\
58107.70 & 176.49 &...&18.80 $\pm$ 0.12&19.33 $\pm$ 0.14&...&18.85 $\pm$ 0.13&18.29 $\pm$ 0.20&LT&This work\\
58107.90 & 176.69 &...&...&...&...&...&18.50 $\pm$ 0.03&LT&This work\\
58113.70 & 182.32 &...&18.86 $\pm$ 0.18&...&...&...&...&LT&This work\\
58123.45 & 191.78 &...&18.68 $\pm$ 0.14&...&18.80 $\pm$ 0.18&18.58 $\pm$ 0.33&...&KeplerCam&Hosseinzadeh et al. 2022\\
58128.70 & 196.87 &...&18.74 $\pm$ 0.09&19.17 $\pm$ 0.14&...&18.70 $\pm$ 0.12&18.22 $\pm$ 0.15&LT&This work\\
58136.51 & 204.45 &...&18.69 $\pm$ 0.09&...&18.66 $\pm$ 0.08&...&...&KeplerCam&Hosseinzadeh et al. 2022\\
58138.09 & 205.98 & 19.99 $\pm$ 0.19&...&19.13 $\pm$ 0.11&...&...&18.02 $\pm$ 0.15&LT&This work\\
58140.51 & 208.33 &...&18.83 $\pm$ 0.11&...&18.77 $\pm$ 0.07&...&...&KeplerCam&Hosseinzadeh et al. 2022\\
58142.12 & 209.89 & 20.10 $\pm$ 0.18&...&19.10 $\pm$ 0.13&...&19.05 $\pm$ 0.12&17.97 $\pm$ 0.14&LT&This work\\
58143.26 & 211.00 &...&18.86 $\pm$ 0.06&...&18.79 $\pm$ 0.05&18.61 $\pm$ 0.04&...&KeplerCam&Hosseinzadeh et al. 2022\\
58144.33 & 212.04 &...&19.00 $\pm$ 0.14&...&18.48 $\pm$ 0.14&...&...&KeplerCam&Hosseinzadeh et al. 2022\\
58145.54 & 213.21 &...&18.77 $\pm$ 0.07&...&18.99 $\pm$ 0.12&...&...&KeplerCam&Hosseinzadeh et al. 2022\\
58146.38 & 214.02 &...&18.93 $\pm$ 0.06&...&18.96 $\pm$ 0.05&18.64 $\pm$ 0.08&...&KeplerCam&Hosseinzadeh et al. 2022\\
58147.43 & 215.04 &...&19.07 $\pm$ 0.16&...&18.98 $\pm$ 0.08&...&...&KeplerCam&Hosseinzadeh et al. 2022\\
58150.26 & 217.79 &...&18.56 $\pm$ 0.12&...&18.86 $\pm$ 0.06&18.49 $\pm$ 0.12&...&KeplerCam&Hosseinzadeh et al. 2022\\
58154.45 & 221.85 &...&18.71 $\pm$ 0.05&...&...&...&...&KeplerCam&Hosseinzadeh et al. 2022\\
58158.46 & 225.75 &...&18.97 $\pm$ 0.06&...&18.87 $\pm$ 0.04&18.32 $\pm$ 0.10&...&KeplerCam&Hosseinzadeh et al. 2022\\
58159.48 & 226.74 &...&19.17 $\pm$ 0.21&...&18.92 $\pm$ 0.07&18.45 $\pm$ 0.07&...&KeplerCam&Hosseinzadeh et al. 2022\\
58164.12 & 231.24 &...&...&...&...&...&18.44 $\pm$ 0.11&AF&This work\\
58164.90 & 231.99 &...&...&...&...&...&18.49 $\pm$ 0.05&LT&This work\\
58166.50 & 233.55 &...&19.12 $\pm$ 0.14&...&19.55 $\pm$ 0.21&...&...&KeplerCam&Hosseinzadeh et al. 2022\\
58171.46 & 238.36 &...&19.79 $\pm$ 0.06&...&...&...&...&KeplerCam&Hosseinzadeh et al. 2022\\
58173.34 & 240.18 &...&19.07 $\pm$ 0.08&...&...&...&...&KeplerCam&Hosseinzadeh et al. 2022\\
58179.35 & 246.01 & ...&...&...&...&19.53 $\pm$ 0.67&...&2.0m LCOGT&This work\\
58182.42 & 248.99 &...&20.43 $\pm$ 0.23&...&20.39 $\pm$ 0.30 &...&...&KeplerCam&Hosseinzadeh et al. 2022\\
58187.44 & 253.86 &...&20.14 $\pm$ 0.12&...&...&19.80 $\pm$ 0.25&...&KeplerCam&Hosseinzadeh et al. 2022\\
58199.96 & 266.01 &$\textgreater$ 21.74 &...&...&...&...&...&LT&This work\\
58211.95 & 277.65 &...&...&...&...&$\textgreater$ 19.97&...&LT&This work\\
58213.37 & 279.02 &...&20.72 $\pm$ 0.27&...&21.63 $\pm$ 0.32&$\textgreater$ 20.96&...&KeplerCam&Hosseinzadeh et al. 2022\\
58218.46 & 283.96 &...&...&...&...&...&19.29 $\pm$ 0.51&LT&This work\\
58221.20 & 286.62 &...&$\textgreater$ 20.73&20.55 $\pm$ 0.21&...&...&...&PO&This work\\
58238.90 & 303.79 &...&...&...&...&...&$\textgreater$ 21.45&AF&This work\\
58293.20 & 356.48 & ...&$\textgreater$ 21.48&$\textgreater$ 20.70&...&...&...&PO&This work\\
\enddata      
\end{deluxetable*}


\begin{deluxetable*}{c c c c c c c c c c}
\tablenum{1}

\tablecaption{\it Continued}
\tablehead{ MJD & Phase$^{a}$ & J & & H & & K &    & Telescope$^{b}$ & Source \\
& (days) & (mag) &  & (mag) &  & (mag) &  & /Inst. &}
\startdata
57906.94 & -18.30 & 16.06 $\pm$ 0.06 &&16.07 $\pm$ 0.06&&16.03 $\pm$ 0.06&&NC& Bose et al. 2018\\
57924.96 & -0.82 & 14.87 $\pm$ 0.06 &&14.92 $\pm$ 0.06&&15.03 $\pm$ 0.06&&NC& Bose et al. 2018\\
57937.90 & 11.74 & 14.86 $\pm$ 0.06 &&14.88 $\pm$ 0.06&&14.91 $\pm$ 0.06&&NC& Bose et al. 2018\\
58043.24 & 113.95 & 16.13 $\pm$ 0.07 &&15.82 $\pm$ 0.06&&15.72 $\pm$ 0.06&&NC& This work\\
58122.26 & 190.62 & 17.72 $\pm$ 0.21 &&17.38 $\pm$ 0.08&&17.13 $\pm$ 0.07&&NC& This work\\
58195.62 & 261.80 & 19.23 $\pm$ 0.20 &&18.78 $\pm$ 0.14&&...&&UKIRT& This work\\
\hline
MJD & Phase$^{a}$ & UVU & UVW1 & UVM2 & UVW2 & & & Telescope$^{b}$ & Source \\
& (days) & (mag) & (mag) & (mag) & (mag) & & & /Inst. &\\
\hline
57906.32 & -18.90 &15.48 $\pm$ 0.05&15.60 $\pm$ 0.03&15.67 $\pm$ 0.03&15.94 $\pm$ 0.03& & &UVOT&Bose et al. 2018\\
57908.69 & -16.60 &15.38 $\pm$ 0.05&15.56 $\pm$ 0.03&15.66 $\pm$ 0.03&15.96 $\pm$ 0.03& & &UVOT&Bose et al. 2018\\
57912.07 & -13.32 &15.18 $\pm$ 0.05&...&...&15.82 $\pm$ 0.03& & &UVOT&Bose et al. 2018\\
57912.41 & -12.99 &15.16 $\pm$ 0.05&...&...&15.79 $\pm$ 0.03& & &UVOT&Bose et al. 2018\\
57912.75 & -12.66 &15.09 $\pm$ 0.05&...&...&15.76 $\pm$ 0.03& & &UVOT&Bose et al. 2018\\
57914.53 & -10.94 &14.94 $\pm$ 0.06&15.22 $\pm$ 0.03&15.33 $\pm$ 0.03&15.68 $\pm$ 0.03& & &UVOT&Bose et al. 2018\\
57915.96 & -9.55 &14.87 $\pm$ 0.05&15.17 $\pm$ 0.03&15.34 $\pm$ 0.03&15.65 $\pm$ 0.03& & &UVOT&Bose et al. 2018\\
57919.21 & -6.39 &14.65 $\pm$ 0.05&15.00 $\pm$ 0.03&15.24 $\pm$ 0.03&15.60 $\pm$ 0.03& & &UVOT&Bose et al. 2018\\
57920.27 & -5.37 &14.58 $\pm$ 0.05&14.88 $\pm$ 0.03&15.11 $\pm$ 0.03&15.46 $\pm$ 0.03& & &UVOT&Bose et al. 2018\\
57922.20 & -3.49 &14.47 $\pm$ 0.05&14.82 $\pm$ 0.03&15.05 $\pm$ 0.03&15.46 $\pm$ 0.03& & &UVOT&Bose et al. 2018\\
57924.20 & -1.55 &14.44 $\pm$ 0.05&14.86 $\pm$ 0.03&15.17 $\pm$ 0.03&15.57 $\pm$ 0.03& & &UVOT&Bose et al. 2018\\
57925.66 & -0.14 &...&...&...&15.68 $\pm$ 0.03& & &UVOT&Bose et al. 2018\\
57926.71 & 0.88 &14.47 $\pm$ 0.05&14.95 $\pm$ 0.03&15.30 $\pm$ 0.03&15.75 $\pm$ 0.03& & &UVOT&Bose et al. 2018\\
57928.71 & 2.82 &14.57 $\pm$ 0.05&15.10 $\pm$ 0.03&15.46 $\pm$ 0.03&15.93 $\pm$ 0.03& & &UVOT&Bose et al. 2018\\
57932.60 & 6.60 &14.64 $\pm$ 0.05&15.36 $\pm$ 0.03&15.84 $\pm$ 0.03&16.27 $\pm$ 0.03& & &UVOT&Bose et al. 2018\\
57934.75 & 8.68 &14.72 $\pm$ 0.05&15.49 $\pm$ 0.03&16.01 $\pm$ 0.03&16.43 $\pm$ 0.03& & &UVOT&Bose et al. 2018\\
57937.37 & 11.23 &14.80 $\pm$ 0.05&15.63 $\pm$ 0.03&16.21 $\pm$ 0.03&16.54 $\pm$ 0.03& & &UVOT&Bose et al. 2018\\
57938.14 & 11.97 &14.84 $\pm$ 0.05&15.70 $\pm$ 0.03&16.26 $\pm$ 0.03&16.62 $\pm$ 0.03& & &UVOT&Bose et al. 2018\\
58026.34 & 97.55 &17.79 $\pm$ 0.12&19.49 $\pm$ 0.15&20.43 $\pm$ 0.27&$\textgreater$ 21.60& & &UVOT&This work\\
58031.38 & 102.44 &18.01 $\pm$ 0.11&19.55 $\pm$ 0.14&20.76 $\pm$ 0.30&21.79 $\pm$ 0.67& & &UVOT&This work\\
58036.33 & 107.25 &18.17 $\pm$ 0.12&20.31 $\pm$ 0.27&...&...& & &UVOT&This work\\
58041.41 & 112.17 &18.34 $\pm$ 0.14&20.17 $\pm$ 0.25&...&...& & &UVOT&This work\\
58046.32 & 116.94 &18.22 $\pm$ 0.14&20.08 $\pm$ 0.27&20.66 $\pm$ 0.40&21.17 $\pm$ 0.47& & &UVOT&This work\\
58051.43 & 121.90 &18.22 $\pm$ 0.12&20.32 $\pm$ 0.26&20.42 $\pm$ 0.25&21.24 $\pm$ 0.41& & &UVOT&This work\\
58056.02 & 126.35 &18.46 $\pm$ 0.16&...&20.51 $\pm$ 0.33&21.31 $\pm$ 0.48& & &UVOT&This work\\
58058.91 & 129.15 &19.07 $\pm$ 0.29&21.48 $\pm$ 0.86&...&...& & &UVOT&This work\\
58083.36 & 152.88 &...&$\textgreater$ 20.90 &$\textgreater$ 21.34&$\textgreater$ 21.47& & &UVOT&This work\\
58096.30 & 165.43 &$\textgreater$ 20.42&...&...&...& & &UVOT&This work\\
\enddata
\tablecomments{\\ 
$\rm ^{a}$ Rest-frame time (days) relative to the epoch of the $g$-band peak brightness.\\
$\rm ^{b}$ The abbreviations of telescope/instrument used are as follows: LT$\rm -- $2.0\,m Liverpool Telescope; PO$\rm -- $0.6\,m telescopes of Post Observatory; LCOGT$\rm -- $Las Cumbres Observatory Global Telescope Network; AF$\rm -- $ALFOSC mounted on 2.0\,m NOT telescope; KeplerCam$\rm -- $KeplerCam mounted on 1.2\,m telescope at Fred Lawrence Whipple Observatory; NC$\rm -- $NotCAM IR imager on 2.0\,m NOT telescope; UKIRT$\rm --$NIR Wide-Field Camera mounted on the United Kingdom Infrared Telescope; UVOT$\rm -- $Ultraviolet Optical Telescope onboard \textit{Swift} satellite.\\
$\rm ^{c}$ We have reprocessed the data with the source of Bose et al. (2018) using the same procedure with this work.\\
$\rm ^{d}$ Swift/UVOT and optical griz photometric data were calibrated for the AB magnitude system while BVJHK for the Vega magnitude system.
}
\end{deluxetable*}

\begin{deluxetable*}{c c c c c c}
\tablenum{2}

\tablecaption{Summary of Late-time Spectroscopic Observations of SN~2017egm.}
\tablehead{ UT date & JD $-$	& Phase$^{a}$ & Exptime & Airmass & Telescope \\
       & 2,458,000 & (days) & (seconds) &  & /Instrument  }
\startdata
2017-09-16.50 & 13.00 & 84.0 & 900 & 2.9 & P200/DBSP \\
2017-09-27.54 & 24.04 & 94.8 & 2100 & 1.9 & Shane/Kast \\
2017-10-08.50 & 35.00 & 105.5 & 720 & 1.5 & Shane/AeroSpOpIR \\
2017-10-19.55 & 46.05 & 116.2 & 1200 & 1.3 & Shane/Kast \\
2017-10-25.54 & 52.04 & 122.0 & 2100 & 1.3 & Shane/Kast \\
2017-10-30.47 & 56.97 & 126.8 & 900 & 1.3 & P200/DBSP \\
2017-10-30.54 & 57.04 & 126.9 & 3000 & 1.2 & Shane/Kast \\
2017-11-16.58 & 74.08 & 143.4 & 2965 & 1.3 & IRTF/SpeX \\
2017-11-21.50 & 79.00 & 148.2 & 1980 & 1.1 & Shane/Kast \\
2017-12-12.48 & 99.98 & 169.0 & 3600 & 1.1 & Shane/Kast \\
2017-12-18.50 & 106.00 & 174.4 & 3600 & 1.0 & Shane/Kast \\
2017-12-28.43 & 115.93 & 184.0 & 3$\times$1000 & 1.1 & P200/DBSP \\
2018-01-13.50 & 132.00 & 199.6 & 3600 & 1.0 & Shane/Kast \\
2018-03-14.29 & 191.79 & 257.6 & 600 & 1.0 & LBT/MODS \\
2018-03-17.99 & 195.49 & 261.2 & 2$\times$1800 & 1.1 & GTC/OSIRIS \\
\enddata
\tablecomments{\\ 
$\rm ^{a}$ Rest-frame time (days) relative to the epoch of the $g$-band peak brightness.\\
}
\label{spec_table}
\end{deluxetable*}

\clearpage

\begin{deluxetable}{r c c c}[htb]

\tablenum{3}
\tablecaption{Best-Fit Blackbody Parameters.\label{bbfit_table}}
\tablehead{	 Phase$^{a}$ & Temperature $T_{\rm bb}$   & Radius $R_{\rm bb}$     &Luminosity $L_{\rm bb}$ \\
 (days)     & ($\rm10^3K$) & ($\rm10^{15}cm$) & ($\rm Log_{10}(erg\ s^{-1})$) }
\startdata
-19.02 & 17.36 $\pm$ 0.15 & 1.18 $\pm$ 0.01 & 43.95 $\pm$ 0.03 \\
-18.05 & 17.03 $\pm$ 0.13 & 1.22 $\pm$ 0.01 & 43.95 $\pm$ 0.02 \\
-17.08 & 16.45 $\pm$ 0.14 & 1.32 $\pm$ 0.02 & 43.96 $\pm$ 0.03 \\
-16.11 & 16.24 $\pm$ 0.11 & 1.37 $\pm$ 0.01 & 43.97 $\pm$ 0.03 \\
-15.39 & 16.13 $\pm$ 0.12 & 1.43 $\pm$ 0.02 & 43.99 $\pm$ 0.02 \\
-10.29 & 15.29 $\pm$ 0.10 & 1.79 $\pm$ 0.02 & 44.10 $\pm$ 0.02 \\
-8.35 & 14.86 $\pm$ 0.09 & 1.95 $\pm$ 0.02 & 44.12 $\pm$ 0.02 \\
-7.38 & 14.53 $\pm$ 0.09 & 2.09 $\pm$ 0.02 & 44.14 $\pm$ 0.02 \\
-6.40 & 14.39 $\pm$ 0.08 & 2.17 $\pm$ 0.02 & 44.16 $\pm$ 0.02 \\
-5.42 & 14.77 $\pm$ 0.10 & 2.15 $\pm$ 0.03 & 44.20 $\pm$ 0.02 \\
-4.45 & 14.53 $\pm$ 0.09 & 2.26 $\pm$ 0.03 & 44.21 $\pm$ 0.02 \\
-1.55 & 13.63 $\pm$ 0.09 & 2.55 $\pm$ 0.03 & 44.20 $\pm$ 0.02 \\
-0.57 & 13.40 $\pm$ 0.07 & 2.61 $\pm$ 0.02 & 44.19 $\pm$ 0.02 \\
0.40 & 12.95 $\pm$ 0.06 & 2.70 $\pm$ 0.02 & 44.17 $\pm$ 0.01 \\
1.37 & 12.73 $\pm$ 0.06 & 2.79 $\pm$ 0.03 & 44.16 $\pm$ 0.02 \\
2.34 & 12.68 $\pm$ 0.06 & 2.71 $\pm$ 0.02 & 44.13 $\pm$ 0.01 \\
3.31 & 12.12 $\pm$ 0.09 & 2.95 $\pm$ 0.05 & 44.13 $\pm$ 0.02 \\
4.28 & 12.03 $\pm$ 0.06 & 2.91 $\pm$ 0.03 & 44.10 $\pm$ 0.02 \\
5.26 & 11.80 $\pm$ 0.05 & 2.95 $\pm$ 0.03 & 44.08 $\pm$ 0.02 \\
6.22 & 11.63 $\pm$ 0.06 & 2.94 $\pm$ 0.03 & 44.05 $\pm$ 0.02 \\
7.19 & 11.34 $\pm$ 0.06 & 3.05 $\pm$ 0.03 & 44.04 $\pm$ 0.02 \\
8.16 & 11.18 $\pm$ 0.05 & 3.07 $\pm$ 0.03 & 44.02 $\pm$ 0.02 \\
9.13 & 11.02 $\pm$ 0.06 & 3.11 $\pm$ 0.03 & 44.01 $\pm$ 0.02 \\
10.10 & 10.83 $\pm$ 0.05 & 3.19 $\pm$ 0.03 & 44.00 $\pm$ 0.02 \\
11.07 & 10.79 $\pm$ 0.06 & 3.14 $\pm$ 0.03 & 43.98 $\pm$ 0.02 \\
11.97 & 10.58 $\pm$ 0.06 & 3.22 $\pm$ 0.03 & 43.97 $\pm$ 0.01 \\
12.04 & 10.57 $\pm$ 0.08 & 3.22 $\pm$ 0.05 & 43.96 $\pm$ 0.03 \\
13.98 & 10.07 $\pm$ 0.18 & 3.42 $\pm$ 0.08 & 43.93 $\pm$ 0.06 \\
15.91 & 10.24 $\pm$ 0.14 & 3.29 $\pm$ 0.08 & 43.92 $\pm$ 0.06 \\
17.85 & 9.88 $\pm$ 0.13 & 3.38 $\pm$ 0.08 & 43.89 $\pm$ 0.05 \\
19.79 & 9.86 $\pm$ 0.12 & 3.41 $\pm$ 0.07 & 43.89 $\pm$ 0.04 \\
20.76 & 9.73 $\pm$ 0.11 & 3.49 $\pm$ 0.07 & 43.88 $\pm$ 0.05 \\
21.73 & 9.72 $\pm$ 0.10 & 3.44 $\pm$ 0.06 & 43.87 $\pm$ 0.04 \\
24.65 & 9.57 $\pm$ 0.08 & 3.47 $\pm$ 0.05 & 43.86 $\pm$ 0.04 \\
27.56 & 9.41 $\pm$ 0.08 & 3.48 $\pm$ 0.10 & 43.83 $\pm$ 0.06 \\
28.53 & 9.34 $\pm$ 0.07 & 3.46 $\pm$ 0.12 & 43.81 $\pm$ 0.07 \\
29.50 & 8.94 $\pm$ 0.12 & 3.70 $\pm$ 0.14 & 43.79 $\pm$ 0.09 \\
99.65 & 6.35 $\pm$ 0.10 & 3.51 $\pm$ 0.12 & 43.15 $\pm$ 0.04 \\
103.55 & 5.97 $\pm$ 0.08 & 3.98 $\pm$ 0.14 & 43.16 $\pm$ 0.05 \\
106.83 & 6.15 $\pm$ 0.13 & 3.39 $\pm$ 0.17 & 43.07 $\pm$ 0.07 \\
111.30 & 6.11 $\pm$ 0.14 & 3.33 $\pm$ 0.19 & 43.04 $\pm$ 0.07 \\
115.17 & 6.58 $\pm$ 0.13 & 2.78 $\pm$ 0.12 & 43.01 $\pm$ 0.04 \\
121.96 & 6.20 $\pm$ 0.11 & 3.17 $\pm$ 0.13 & 43.03 $\pm$ 0.05 \\
126.82 & 6.91 $\pm$ 0.10 & 2.38 $\pm$ 0.07 & 42.96 $\pm$ 0.03 \\
128.75 & 6.59 $\pm$ 0.14 & 2.34 $\pm$ 0.11 & 42.87 $\pm$ 0.04 \\
135.50 & 5.78 $\pm$ 0.34 & 2.17 $\pm$ 0.30 & 42.57 $\pm$ 0.16 \\
137.45 & 6.45 $\pm$ 0.70 & 1.54 $\pm$ 0.35 & 42.47 $\pm$ 0.19 \\
\enddata
\end{deluxetable}

\begin{deluxetable}{r c c c}[ht]
\tablenum{3}
  \tablecaption{\it Continued}

\tablehead{	 Phase$^{a}$ & Temperature $T_{\rm bb}$   & Radius $R_{\rm bb}$     &Luminosity $L_{\rm bb}$ \\
 (days)     & ($\rm10^3K$) & ($\rm10^{15}cm$) & ($\rm Log_{10}(erg\ s^{-1})$) }
\startdata
139.35 & 5.72 $\pm$ 0.51 & 1.87 $\pm$ 0.39 & 42.43 $\pm$ 0.22 \\
140.41 & 5.81 $\pm$ 0.35 & 1.69 $\pm$ 0.22 & 42.36 $\pm$ 0.13 \\
143.41 & 5.07 $\pm$ 0.35 & 1.92 $\pm$ 0.28 & 42.24 $\pm$ 0.12 \\
144.38 & 5.36 $\pm$ 0.34 & 1.72 $\pm$ 0.22 & 42.24 $\pm$ 0.11 \\
145.35 & 5.22 $\pm$ 0.42 & 1.68 $\pm$ 0.32 & 42.17 $\pm$ 0.20 \\
148.26 & 4.78 $\pm$ 0.39 & 1.95 $\pm$ 0.40 & 42.15 $\pm$ 0.26 \\
151.17 & 5.17 $\pm$ 0.55 & 1.61 $\pm$ 0.39 & 42.12 $\pm$ 0.22 \\
152.13 & 6.30 $\pm$ 1.37 & 0.99 $\pm$ 0.43 & 42.04 $\pm$ 0.28 \\
159.85 & 6.02 $\pm$ 1.23 & 1.42 $\pm$ 0.59 & 42.28 $\pm$ 0.29 \\
165.63 & 7.19 $\pm$ 0.63 & 1.03 $\pm$ 0.17 & 42.31 $\pm$ 0.13 \\
167.57 & 7.85 $\pm$ 1.06 & 0.82 $\pm$ 0.20 & 42.26 $\pm$ 0.21 \\
168.54 & 7.44 $\pm$ 1.02 & 0.91 $\pm$ 0.22 & 42.26 $\pm$ 0.18 \\
171.45 & 7.89 $\pm$ 0.93 & 0.82 $\pm$ 0.16 & 42.27 $\pm$ 0.21 \\
204.45 & 9.46 $\pm$ 1.02 & 0.62 $\pm$ 0.11 & 42.34 $\pm$ 0.26 \\
208.33 & 9.20 $\pm$ 1.06 & 0.63 $\pm$ 0.12 & 42.31 $\pm$ 0.28 \\
211.00 & 7.32 $\pm$ 0.40 & 0.99 $\pm$ 0.09 & 42.30 $\pm$ 0.10 \\
212.04 & 6.34 $\pm$ 0.48 & 1.26 $\pm$ 0.17 & 42.26 $\pm$ 0.12 \\
213.21 & 7.96 $\pm$ 0.65 & 0.85 $\pm$ 0.11 & 42.31 $\pm$ 0.16 \\
214.02 & 8.47 $\pm$ 0.76 & 0.72 $\pm$ 0.11 & 42.28 $\pm$ 0.17 \\
215.04 & 5.58 $\pm$ 0.53 & 1.54 $\pm$ 0.30 & 42.22 $\pm$ 0.11 \\
217.79 & 7.22 $\pm$ 1.40 & 0.96 $\pm$ 0.30 & 42.25 $\pm$ 0.28 \\
225.75 & 7.67 $\pm$ 0.66 & 0.86 $\pm$ 0.14 & 42.26 $\pm$ 0.12 \\
233.55 & 6.90 $\pm$ 1.36 & 0.89 $\pm$ 0.32 & 42.11 $\pm$ 0.23 \\
248.99 & 7.19 $\pm$ 1.45 & 0.59 $\pm$ 0.26 & 41.83 $\pm$ 0.28 \\
\enddata
\tablecomments{\\ 
$\rm ^{a}$ Rest-frame time (days) relative to the epoch of the $g$-band peak brightness.\\
Data after +148\,d were fitted without UV limit.
}
\end{deluxetable}

\begin{deluxetable*}{c c c c c c c}[hb]

\tablenum{4}
  \tablecaption{Blackbody Fitting Results for Epochs with MIR Data}\label{IR_bb_table}

\tablehead{	 Phase$^{a}$ & $T_{\rm IR,\ bb}$ &  $R_{\rm IR,\ bb}$ &  $L_{\rm IR,\ bb}$ &  $T_{\rm Opt,\ bb}$  & $R_{\rm Opt,\ bb}$ & $L_{\rm Opt,\ bb}$  \\
 (days)  & ($\rm K$) & ($\rm10^{15}cm$) & $\rm (10^{41}erg\ s^{-1})$ & ($\rm K$) &($\rm10^{15}cm$) & $\rm (10^{41}erg\ s^{-1})$ }
\startdata
+137 & 520 $\pm$ 70  & 82.1 $\pm$ 12.0 & 3.54 $\pm$ 1.50 & 7100 $\pm$ 500 & 1.2 $\pm$ 0.2 & 25.30 $\pm$ 9.70 \\
+292 & 565 $\pm$ 83  & 73.1 $\pm$ 14.7 & 3.89 $\pm$ 2.01 & 4800 $\pm$ 420 & 1.5 $\pm$ 0.5 & 5.51 $\pm$ 2.08 \\
+492 & 827 $\pm$ 45  & 22.2 $\pm$ 2.5 & 1.65 $\pm$ 0.40 & $-$ & $-$ & $-$\\
\enddata

\end{deluxetable*}

\end{document}